\documentclass[12pt,preprint]{aastex}

\slugcomment{}
\shorttitle{Flare-ona of EK Dra}
\shortauthors{T.\ R.\ Ayres}

\begin{document}

\title{The Flare-ona of EK Draconis}

\author{Thomas R.\ Ayres}

\affil{Center for Astrophysics and Space Astronomy,\\
389~UCB, University of Colorado,
Boulder, CO 80309;\\ Thomas.Ayres@Colorado.edu}

\begin{abstract}

EK~Draconis (HD\,129333: G1.5~V) is a well-known young (50~Myr) solar analog.  In 2012, {\em Hubble Space Telescope}\/ returned to EK~Dra to follow up a far-ultraviolet (FUV) SNAPshot visit by Cosmic Origins Spectrograph (COS) two years earlier.  The brief SNAP pointing had found surprisingly redshifted, impulsively variable subcoronal ``hot-line'' emission of \ion{Si}{4} 1400~\AA\ ($T\sim 8\times10^4$~K).  Serendipitously, the 2012 follow-on program witnessed one of the largest FUV flares ever recorded on a sunlike star, which again displayed strong redshifts (downflows) of 30--40~km s$^{-1}$, even after compensating for small systematics in the COS velocity scales, uncovered through a cross-calibration by Space Telescope Imaging Spectrograph (STIS).  The (now reduced, but still substantial) $\sim$10~km s$^{-1}$ hot-line redshifts outside the flaring interval did not vary with rotational phase, so cannot be caused by ``Doppler Imaging'' (bright surface patches near a receding limb).  Density diagnostic \ion{O}{4}] 1400~\AA\ multiplet line ratios of EK~Dra suggest $n_{\rm e}\sim 10^{11}$ cm$^{-3}$, an order of magnitude larger than in low-activity solar twin $\alpha$~Centauri A, but typical of densities inferred in large stellar soft X-ray events.  The self-similar FUV hot-line profiles between the flare decay and the subsequent more quiet periods, and the unchanging but high densities, reinforce a long-standing idea that the coronae of hyperactive dwarfs are flaring all the time, in a scale-free way; a {\em flare-ona}\/ if you will.  In this picture, the subsonic hot-line downflows probably are a byproduct of the post-flare cooling process, something like ``coronal rain'' on the Sun.  All in all, the new STIS/COS program documents a complex, energetic, dynamic outer atmosphere of the young sunlike star.

\end{abstract}


\section{INTRODUCTION}

Flares on the Sun result from catastrophic releases of energy previously stored in strong tangled surface magnetic fields.  These outbursts are most conspicuous at high energies, and under the right circumstances (or wrong, depending on your perspective) can profoundly impact the heliosphere and Earth.  This is one reason why flares have drawn intense interest from solar observers and theorists for many decades; and indeed have spawned a vibrant Space Weather Community in recent years.  A less appreciated aspect of solar flares involves the stellar cousins, where extraordinarily large X-ray (and even white-light) outbursts have been recorded, in some cases dwarfing by a considerable margin the largest known flares on our own Sun, albeit a middle-aged and magnetically rather uninspired star.  The most extreme of these are ``superflares\footnote{Originally called ``flashes'' by Schaefer (1989), who first called attention to the class.},'' which at high energies can exceed the quiescent coronal X-ray emission of a star by factors of a hundred or more.  Interest in such rare, but potentially catastrophic, events concerns how they might affect, for example, survival of primitive lifeforms struggling for existence on a semi-habitable world orbiting a young, magnetically active star.

Until recently, reports of superflares mainly involved interpretations of historical materials, such as long-term astronomical plate collections, and disconnected isolated reports or inferences (Schaefer et al.\ 2000).  Now, however, the {\em Kepler}\/ Mission has provided more substantive statistics on superflares -- although limited to white-light counterparts -- from a wide variety of stars, normal and otherwise (e.g., Maehara et al.\ 2012; Shibayama et al.\ 2013; Candelaresi et al.\ 2014).  Further, the {\em Swift}\/ gamma-ray burst satellite has witnessed a handful of the most extreme events, radiating in hard X-rays, thanks to its broad view of the high-energy sky and rapid pointing response (e.g., Osten et al.\ 2010; Drake et al.\ 2014).  The {\em Swift}\/ alerts also have promoted supporting multi-spectral observations of flares, especially at radio frequencies that capture non-thermal emissions from accelerated particles (e.g., Fender et al.\ 2015).  On the other hand, there have been few cases of such outbursts recorded at wavelengths where the powerful diagnostics of high-resolution spectroscopy can be brought to bear, namely the ultraviolet.  The lack of good examples mostly is because these rare events are, well, rare.

Serendipitously, however, a recent {\em HST}\/ Cosmic Origins Spectrograph (COS) campaign on young ($\sim$50~Myr) solar analog EK Draconis (HD\,129333; G1.5~V), well-described in the previous literature (e.g., K{\"o}nig et al.\ 2005, and references therein), caught a large hour-long FUV outburst, in ``hot lines'' like the \ion{C}{4} 1550~\AA\ doublet (T$\sim 1\times 10^{5}$~K) and the extremely hot [\ion{Fe}{21}] 1354~\AA\ coronal forbidden line ($\sim 10^{7}$~K).  The EK~Dra flare, viewed with the excellent spectral resolution, broad wavelength coverage, and high sensitivity of COS, provides a unique opportunity to explore properties of such events, to inform interpretations of the physical processes involved.  Moreover, the behavior of the FUV spectrum outside the flare, over the several days of intermittent attention by {\em HST,}\/ opens a window onto the ``quiescent'' properties of a hyperactive chromosphere and corona, again with unprecedented clarity.

\section{OBSERVATIONS}

\subsection{Previous {\em HST}\/ Studies}

There were two previous {\em HST}\/ spectroscopic programs of note on EK~Dra.  In 1996, Saar \& Bookbinder (1998) carried out high time resolution measurements with the low-resolution FUV grating of Goddard High-Resolution Spectrograph (GHRS) over a continuous interval of about six hours (about 10\% of the spin period).  The authors identified several factor of 2-3 impulsive flare-like bursts in \ion{Si}{4} 1400~\AA\ and \ion{C}{4} 1550~\AA\ when the light curves were binned into 30~s intervals.  The identifiable hot-line emission enhancements occupied about 8\% of the timeline.  The remainder was attributed to ``quiescent'' periods.

A decade and a half later, in 2010, Ayres \& France (2010) targeted EK~Dra with newly commissioned COS as part of a larger ``SNAPshot'' (partial-orbit filler observations) survey of FUV coronal forbidden lines, such as [\ion{Fe}{21}] $\lambda$1354 ($T\sim 10^{7}$~K), in more than a dozen late-type stars.  The objective was to measure Doppler profiles of high-energy coronal lines at FUV resolution, some 20 times better than currently possible in the soft X-ray band, the more natural -- but spectroscopically challenged -- home of coronal transitions.  The COS SNAP observation of EK~Dra covered only a short time span, about 20~minutes, compared with the earlier GHRS pointing.  But, it had the important advantage of achieving moderately high spectral resolution ($\sim 20$~km s$^{-1}$) in the integrated spectrum, as well as comparably high time resolution (30~s) for flux measurements of key species like \ion{C}{2} 1335~\AA\ ($2{\times}10^{4}$~K), \ion{Si}{4}, and [\ion{Fe}{21}] collected in the single G130M grating setting feasible for the brief partial-orbit visit.

The \ion{Si}{4} doublet, in particular, exhibited twin factor of $\sim$2 enhancements over the mere 20~minute interval, remarkably without comparable responses at lower temperatures (\ion{C}{2}) or higher (\ion{Fe}{21}), contrary to expectations for normal flares, which tend to impact all the atmospheric temperature layers more-or-less simultaneously.  Equally remarkable, the \ion{Si}{4} doublet features displayed quite large {\em redshifts}, about 20~km s$^{-1}$, relative to narrow chromospheric features (which set the empirical zero-point velocity for the COS exposures).

Redshifts of the hot lines are well documented on the Sun, attributed to nearly radial downflows of subcoronal material, but the average shifts in a disk-integrated spectrum might be only a few km~s$^{-1}$ or less (Peter 2006).  Ayres \& France interpreted the large redshifts of \ion{Si}{4} on EK~Dra as a signature of a scaled up version of solar ``coronal rain:'' previously hot multi-MK gas suspended high in magnetic loops rapidly cooling and draining back down to the lower atmosphere, perhaps -- in the case of EK~Dra -- in a catastrophic way (what one might call an {\em anti-flare:}\/ impulsive cooling rather than heating).

\subsection{The 2012 {\em HST}\/ Campaign}

The follow-up {\em HST}\/ program, described here, partnered the high sensitivity of COS with the excellent wavelength precision of STIS to explore alternative explanations for the COS redshifts, and thus also the coronal rain hypothesis: (1) subtle distortions in the COS wavelength scales; and (2) ``Doppler Imaging'' effects, whereby active patches on the stellar surface might appear in the integrated spectrum as a redshift if, say, a bright region happened to be near the receding limb (where the projected rotational velocity could be as much as $+\upsilon\sin{i}\sim +17$~km s$^{-1}$).  A secondary objective was to carry out a more extensive time-domain study of FUV variability, along the lines of the Saar \& Bookbinder work, but with the added dimension of time-resolved spectroscopy possible with high-sensitivity, time-tagged COS.  A related goal was to extend the previous minimal SNAP wavelength coverage (G130M/detector side A: 1290-1430~\AA), adding G160M/sides A and B, especially to reach the important \ion{C}{4} 1550~\AA\ doublet and \ion{He}{2} $\lambda$1640 at the longer wavelengths; and with STIS to probe below the COS G130M/A shortwavelength cutoff (side B was deactivated because \ion{H}{1} Ly$\alpha$ $\lambda$1215 violated detector bright limits), including especially Ly$\alpha$ itself.

The characteristics and operational capabilities of COS have been described in a number of previous publications, especially Green et al.\ (2012); and companion UV spectrograph STIS, by Woodgate et al.\ (1998) and Kimble et al.\ (1998).

The new {\em HST}\/ program was carried out over two days in late-March 2012 (the stellar rotation period is 2.8~d).  EK~Dra was in the {\em HST}\/ Continuous Viewing Zone (CVZ) at the time, by design, which allowed each of the six allocated 96~minute orbits to be used fully, without any loss to Earth occultations.  Figure~1 is a schematic timeline of the program, also summarized in Table~1.

\clearpage
\begin{figure}[ht]
\figurenum{1}
\vskip  0mm
\hskip  -13mm
\includegraphics[scale=0.75,angle=90]{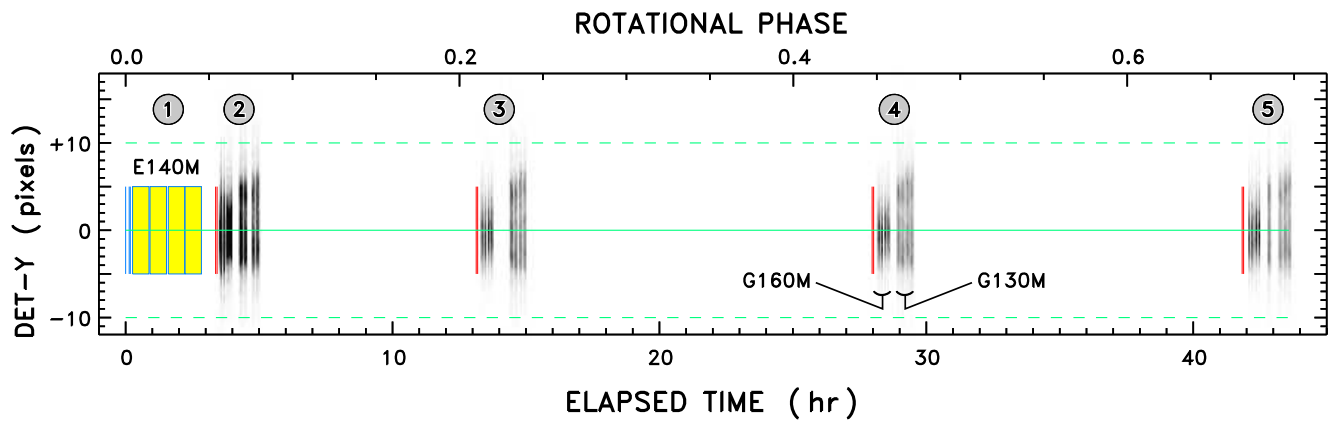} 
\vskip -5mm
\figcaption[]{\small Observational timeline of the 2012 EK~Dra {\em HST}\/ program.  Circled numerals are visit numbers.  Yellow blocks represent initial STIS E140M exposure, split into four equal length sub-exposures; preceded by a target acquisition and peak-up (thin blue vertical lines).  First COS observation (visit 2) followed immediately, with grating setting G160M, then G130M, both split into four FP-POS steps.  Each FP-POS sub-exposure is represented by a streak image derived from the associated event list (darker shading represents higher count rates).  Non-uniform spacing of the G130M sub-exposures in visits~2 and 5 devolve from the scheduling process.  Vertical red lines mark the COS NUV imaging target acquisitions. 
}
\end{figure}

\begin{deluxetable}{cccccc}
\tabletypesize{\footnotesize}
\tablenum{1}
\tablecaption{{\em HST}\/ Observations of EK~Dra (2010--2014)}
\tablecolumns{6}
\tablewidth{0pt}
\tablehead{\colhead{Dataset} & \colhead{UT Start} & \colhead{$t_{\rm exp}$~(s)} & \colhead{Aperture}  &  \colhead{Grating$-\lambda_{\rm cen}$(\AA)} &  \colhead{FP-POS}\\
\colhead{(1)} & \colhead{(2)} & \colhead{(3)} & \colhead{(4)} & \colhead{(5)} & \colhead{(6)}
} 
\startdata
\cutinhead{2010 SNAP Program}
lb3e34010 & Apr-22~09:18  & 1160   & PSA   & G130M-1291 & 3,4  \\[-3pt] 
\cutinhead{2012 Visits 1 \& 2 (contiguous)}
oboq01010 & Mar-27~16:41  & 8880   & PHT   & E140M-1425 & \nodata \\[-3pt]
lboq02010 & Mar-27~19:57  & 1000   & PSA   & G160M-1577 & 3,4,1,2 \\[-3pt]    
lboq02020 & Mar-27~20:41  & 1780   & PSA   & G130M-1291 & 3,4,1,2 \\[-3pt]
\cutinhead{2012 Visit 3}
lboq03010 & Mar-28~05:43  & 1000   & PSA   & G160M-1577 & 3,4,1,2 \\[-3pt]
lboq03020 & Mar-28~06:48  & 1780   & PSA   & G130M-1291 & 3,4,1,2 \\[-3pt]
\cutinhead{2012 Visit 4}
lboq04010 & Mar-28~20:33  & 1000   & PSA   & G160M-1577 & 3,4,1,2 \\[-3pt]
lboq04020 & Mar-28~21:16  & 1780   & PSA   & G130M-1291 & 3,4,1,2 \\[-3pt]
\cutinhead{2012 Visit 5}
lboq05010 & Mar-29~10:25  & 1000   & PSA   & G160M-1577 & 3,4,1,2 \\[-3pt]
lboq05020 & Mar-29~11:08  & 1780   & PSA   & G130M-1291 & 3,4,1,2 \\[-3pt]
\cutinhead{2014 SNAP Program}
ocd839010 & May-06~10:10  & ~600   & PHT   & E230H-2713 & \nodata
\enddata
\tablecomments{Col.~1 prefix ``o'' is STIS, ``l'' is COS. Col.~4 ``PHT'' is the STIS $0.2^{\prime\prime}{\times}0.2^{\prime\prime}$ photometric aperture; PSA is the COS $2.5^{\prime\prime}$-diameter Primary Science Aperture.  Col.~6 indicates sequence of (standard) small grating steps to mitigate fixed pattern noise.  The 2014 STIS observation was from an unrelated program (S.~Redfield P.I.), but nonetheless relevant to the present study.}
\end{deluxetable}

\clearpage
The first visit, consisting of two orbits, featured a deep STIS exposure with the medium-resolution FUV echelle (E140M-1425) for 8.8~ks, split into four equal-length sub-exposures (to mitigate instrumental drifts due to telescope ``breathing'').  The initial target acquisition was with the CCD in visible light through the F25ND3 filter.  Following that, a peak-up (accurate target centering via a raster search) was performed with the $0.2^{\prime\prime}{\times}0.06^{\prime\prime}$ slit in dispersed visible light (CCD with G430M-3936 grating setting).  Then, the STIS FUV exposures were taken, through the $0.2^{\prime\prime}{\times}0.2^{\prime\prime}$ photometric aperture for maximum throughput, although in principle retaining high accuracy in the wavelength scales thanks to the narrow-slit peak-up.

Immediately following the STIS segment, although in a separate visit\footnote{COS utilizes different Fine Guidance Sensor (FGS) sky segments (``pickles''), and thus a different set of Guide Stars; so a separate Guide-Star/target acquisition had to be performed.}, was the first of four one-orbit COS pointings, spaced evenly over roughly half a spin cycle so all sides of the possibly inhomogeneous surface activity would be visible.

The four COS visits ostensibly were identical and occupied a single CVZ orbit each.  The COS visits began with an NUV imaging target acquisition, using Mirror-A and the Bright Object Aperture (BOA).  Following that, a 1000~s observation was taken with G160M-1577 (1387-1557~\AA\ on detector side B; 1578-1749~\AA\ on side A), divided into equal length sub-exposures, one at each of the standard four ``FP-POS'' positions (small grating rotations to suppress fixed-pattern noise).  Then, a 1780~s integration, also split into four equal FP-POS sub-exposures, was taken with G130M-1291 (1292-1432~\AA\ on side A; side B turned off, as in the earlier SNAP program, owing to overbright Ly$\alpha$).

All the COS observations were done in TIME-TAG mode, which delivers a time-stamped record of photon $(x,y)$ positions in the 2-D detector format ($x$ corresponds to wavelength; $y$ to cross-dispersion spatial coordinate).  These photon event lists can be processed in a number of ways to provide time histories of fluxes and/or line profiles.  The particular G130M and G160M settings were chosen because they overlap at the important \ion{Si}{4} 1400~\AA\ doublet, so that it would be recorded in all the observations.  Unfortunately, the key [\ion{Fe}{21}] coronal forbidden line was captured only in the G130M segments; and the important subcoronal \ion{C}{4} doublet only in the G160Ms.

\subsubsection{STIS Observations in Visit 1}

The initial STIS observation and first of the COS visits were intended to be carried out as a pair, as close together as practical, to serve as a velocity cross-validation.  Although COS exposures have an internal wavelength reference (lamp flash), the calibration spectrum falls well above the stellar stripe, on a part of the detector where uncompensated geometrical distortions are different than where the stellar spectrum is located.  Further, the specific $y$-position of the stellar spectrum, and thus the local geometrical distortions it experiences, depends on the acquisition method.  However, because the same strategy was used for all four COS visits, any wavelength issues identified in the comparison of the initial COS visit with STIS should apply equally well to the three subsequent COS pointings.  At the same time, while STIS has exquisite velocity precision and can record S/N$\sim$20 profiles of the brightest lines (including Ly$\alpha$, forbidden to COS), it lacks the sensitivity to resolve shapes of the bright lines on short time scales, or dig down to fainter features like the key \ion{O}{4}] 1400~\AA\ density diagnostic (see, e.g., Cook et al.\ 1995, and references therein).

The STIS spectra of EK~Dra were processed and combined using protocols developed for the {\em HST}\/ Advanced Spectral Library Project (ASTRAL\footnote{see: http://casa.colorado.edu/$\sim$ayres/ASTRAL/}), which includes a post-pipeline correction for subtle wavelength distortions.  The four independent sub-exposures were co-added, but without the usual cross-correlation registration, because the relatively faint spectra lacked narrow features of suitable strength for the purpose.  Figure~2 compares representative intervals of the average STIS FUV echelle spectrum of EK~Dra to low-activity solar twin $\alpha$~Centauri A (G2~V), also observed by STIS with the same setup (see Ayres 2015).

At first glance, the two spectra are surprisingly alike, despite the enormous gulf between the stars in coronal activity (the $L_{\rm X}$ of EK~Dra is about a thousand times that of $\alpha$~Cen).  For example, the narrow chromospheric lines of \ion{O}{1}, \ion{C}{1}, and \ion{C{\em l}}{1} are quite similar in appearance between the hyperactive star and low-activity counterpart.  Even the hot lines like \ion{C}{4} have similar emission cores. 

\begin{figure}
\figurenum{2}
\vskip  0mm
\hskip -5mm
\includegraphics[scale=0.70,angle=90]{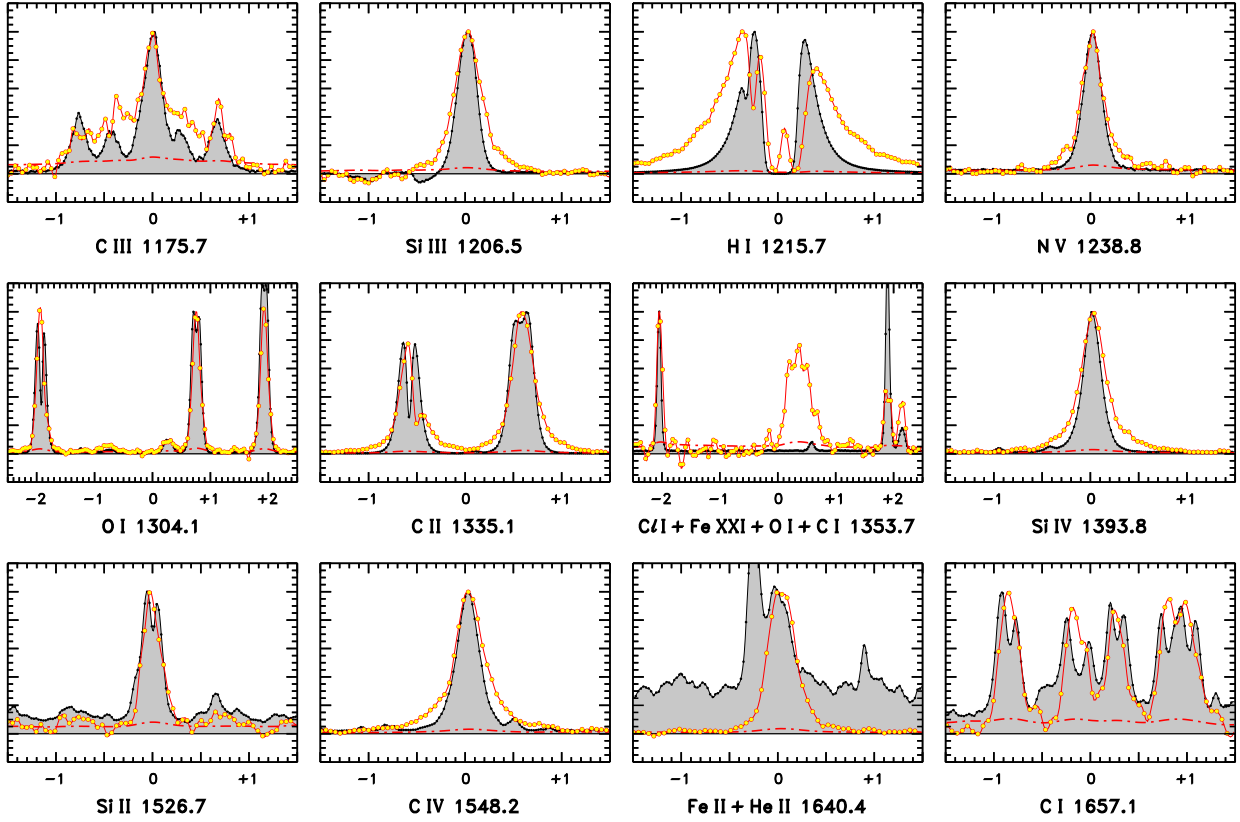} 
\vskip 0mm
\figcaption[]{\small Examples of STIS E140M FUV echelle spectra of EK~Dra (open circles; 4-point smoothing [2 resolution elements (resels)]) compared to low-activity solar analog $\alpha$~Cen A (gray shading outlined by black dots).  The $x$-axis scale in each panel is $\Delta\lambda$ (\AA) relative to the wavelength cited in the axis title.  Red dotted-dashed curves are the smoothed 1\,$\sigma$ photometric noise levels (per resel) for EK~Dra.  The two spectra were adjusted to match the peak of one of the spectral features in each interval.  The wavelength scales were registered to the average velocity shift of narrow chromospheric emissions.  Conspicuous absorptions in several of the chromospheric features (e.g., \ion{H}{1} $\lambda$1215 Ly$\alpha$, \ion{O}{1} $\lambda$1302, and \ion{C}{2} $\lambda$1334) are from the interstellar medium (ISM).  The ISM velocity is more redshifted in EK~Dra, relative to the stellar radial velocity, than in $\alpha$~Cen A, so the EK~Dra absorptions cut into the red edges of the narrower emission cores (e.g., \ion{O}{1} $\lambda$1302 and \ion{C}{2} $\lambda$1334).  The small peak in the center of the EK~Dra Ly$\alpha$ profile is atomic hydrogen skyglow through the $0.2^{\prime\prime}{\times}0.2^{\prime\prime}$ aperture.  The dip in the blue wing of \ion{Si}{3} $\lambda$1206 in $\alpha$~Cen is an artifact due to contamination of the local background by Ly$\alpha$ light scattered from the adjacent echelle order.
}
\end{figure}

Closer examination, however, reveals several differences.  The most noticeable is the exceptionally broad \ion{H}{1} Ly$\alpha$ profile of EK~Dra (although somewhat emphasized by the normalization approach: more distant EK~Dra [$d\sim 34$~pc] is more heavily absorbed by interstellar \ion{H}{1} than close-by $\alpha$~Cen [1.3~pc], so is missing much of its intrinsic narrow peak).  A second, in fact blatant, difference is the bright [\ion{Fe}{21}] coronal forbidden line of EK~Dra, which is absent in low-activity $\alpha$~Cen (the weak, narrow feature at 1354~\AA\ is a chromospheric \ion{C}{1} line).  On the other hand, the striking difference at the \ion{He}{2} $\lambda$1640 Balmer-$\alpha$ line mainly comes about because low-activity $\alpha$~Cen is dominated by a chromospheric \ion{Fe}{2} blend on the blueward side of \ion{He}{2}; rather than high-excitation \ion{He}{2} itself as in EK~Dra.  The prominent photospheric continuum of $\alpha$~Cen A in the \ion{He}{2} $\lambda$1640 panel is an artifact of the scaling procedure: the line-to-continuum contrast is much smaller in the low-activity star compared with hyperactive EK~Dra.  If the normalization instead had been according to the bolometric fluxes, the 1640~\AA\ photospheric continuum levels of the two stars would be similar, but the \ion{He}{2} emission of EK~Dra would dwarf that of $\alpha$~Cen A.  (Indeed, a similar dwarfing would occur in all the panels with an $f_{\rm bol}$ normalization.)  More subtle, but significant, differences are the broad low-intensity wings of the EK~Dra hot lines: \ion{Si}{3}, \ion{Si}{4}, \ion{C}{4}, and \ion{N}{5} ($T\sim 0.5-2\times10^{5}$~K).  There are some additional differences seen in the 1353.7~\AA\ panel, having to do with pumping of the \ion{C{\em l}}{1} and \ion{C}{1} lines by higher excitation species, leaving \ion{O}{1} $\lambda$1355 much weaker by comparison in the active star.

\begin{figure}[ht]
\figurenum{3}
\vskip  0mm
\hskip -10mm
\includegraphics[scale=0.75,angle=90]{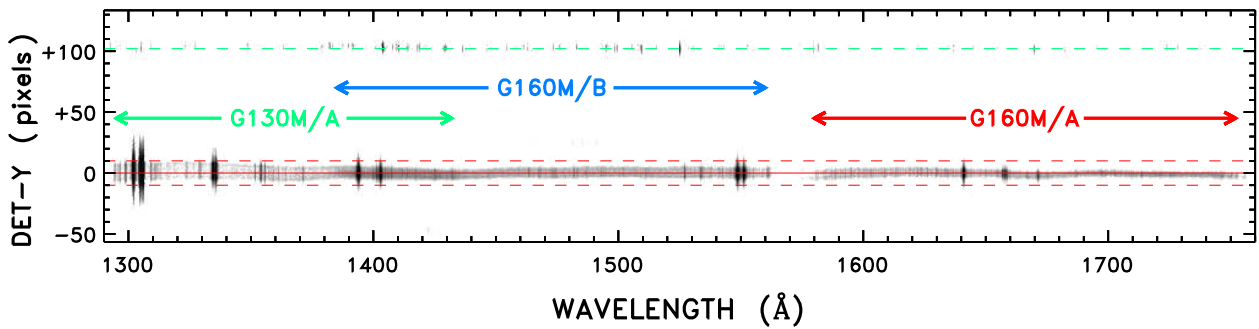} 
\vskip 0mm
\figcaption[]{\small Schematic EK~Dra COS FUV spectrum from the 2012 campaign.  The G130M and G160M settings overlap in the important 1400~\AA\ region.  Note the detector gap between G160M sides A and B at 1560-1580~\AA.  Image was constructed from the totality of the associated photon event lists from COS visits 2-5, with adjustments in the detector $y$ centroids to align the three separate segments, for display purposes.
}
\end{figure}

\subsubsection{COS Observations in Visits 2-5}

Figure~3 is a schematic of the COS FUV spectrum covered by the 2012 program.  The lamp-flash calibrations (barely visible at this intensity stretch) can be seen 100 pixels above the main stellar stripe.  The two G160M segments were observed simultaneously.  The longwavelength end of G130M on detector side A overlaps the shortwavelength portion of G160M on side B at the key \ion{Si}{4} 1400~\AA\ resonance doublet (and \ion{O}{4}] density-diagnostic multiplet); side B of G130M was disallowed, as in the 2010 SNAP program, owing to intense chromospheric Ly$\alpha$.  The broad features (in cross-dispersion $y$ coordinate) at 1300~\AA\ are the \ion{O}{1} resonance triplet, dominated by sky-glow through the 2.5$^{\prime\prime}$-diameter COS aperture.  The other bright features are \ion{C}{2} $\lambda$1335, \ion{Si}{4} $\lambda$1400 (distinct pair), and \ion{C}{4} $\lambda$1550 (close pair).

\begin{deluxetable}{lccccc}
\tabletypesize{\footnotesize}
\tablenum{2}
\tablecaption{Measurement Parameters for the COS Event Lists}
\tablecolumns{6}
\tablewidth{0pt} 
\tablehead{\multicolumn{3}{c}{\underline{~~~~Integrated Count Rates~~~~}}  &  \multicolumn{3}{c}{\underline{~~~~~~~Spectral Profiles~~~~~~~}} \\
\colhead{Type} & \colhead{Band~(\AA)} & \colhead{(S/N)$_{\rm crit}$} & \colhead{$\lambda_{0}$~(\AA)} & \colhead{$\Delta{\lambda}$~(\AA)}  &  \colhead{$N_{\rm crit}$~(counts)}\\
\colhead{(1)} & \colhead{(2)} & \colhead{(3)} & \colhead{(4)} & \colhead{(5)} & \colhead{(6)} 
} 
\startdata
\cutinhead{G130M-1291}
LINE-1 & $1335.1{\pm}1.5$  &  20  &  1335.703  &  0.05  &  3000 \\[3pt]
LINE-2 & $1354.0{\pm}0.75$  &  10  &  1354.080  &  0.10  &  1000 \\[3pt]
LINE-3 & $1393.8{\pm}1.0$\,+\,$1402.8{\pm}0.75$  &  20  &  1402.773  &  0.05  &  1000 \\[3pt]
CON-A & $1344.5{\pm}6.0$\,+\,$1381.0{\pm}8.0$   &  20  &  \nodata  &  \nodata & \nodata \\[3pt]
BKG-A & $1360{\pm}70$   &  \nodata   &  \nodata  &  \nodata & \nodata \\[3pt]
\cutinhead{G160M-1577}
LINE-1 & $1393.8{\pm}1.0$\,+\,$1402.8{\pm}0.75$  &  20  &  1402.773  &  0.05  &  1000 \\[3pt]
LINE-2 & $1549.5{\pm}2.75$  &  20  &  1548.204  &  0.05  &  2000 \\[3pt]
LINE-3 & $1640.35{\pm}1.0$  &  20  &  1640.404  &  0.05  &  1000 \\[3pt]
CON-B & $1435{\pm}25$   &  20  &  \nodata  &  \nodata & \nodata \\[3pt]
BKG-B & $1475{\pm}85$   &  \nodata  &  \nodata  &  \nodata & \nodata \\[3pt]
CON-A & $1610{\pm}25$   &  20  &  \nodata  &  \nodata & \nodata \\[3pt]
BKG-A & $1665{\pm}85$   &  \nodata  &  \nodata  &  \nodata & \nodata
\enddata
\tablecomments{Col.~1: ``LINE'' for emission line, ``CON'' for continuum band, ``BKG'' for off-spectrum background.  Suffixes ``B'' and ``A'' refer to the respective detector segments.  For the integrated count rates, col~2.\ is the integration bandpass, and col.~3 is the signal-to-noise of each measurement (defining the duration, $\Delta{t}$, of the bin).  For the line shape extractions, col.~4 is the central wavelength of the constructed profile (${\pm}250$~km s$^{-1}$ bandpass), col.~5 is the bin size, and col.~6 is the number of counts collected in each profile (again, defining a $\Delta{t}$).  For the fluxes and profiles, the cross-dispersion ($y$) extraction window was ${\pm}9$~pixels.  Background swaths were positioned at $\Delta{y}= {\pm}40$ pixels from the spectrum center, with 20 pixel widths.  Background count rates were determined from the full duration of each subexposure, regardless of S/N.}
\end{deluxetable}

As mentioned earlier, the initial STIS and COS visits were intended to be as close as possible to enable a cross-comparison of the two UV instruments.  Thanks to diligent work by the {\em HST}\/ schedulers, the first COS exposure (G160M) began a mere 40~minutes after the final STIS readout, normally close enough to ensure comparable stellar conditions over the three orbits of contiguous STIS and COS attention.  Indeed, the STIS FUV emission line intensities were stable over the 2.5~hours of near-continuous E140M exposure.  As luck would have it, however, less than an hour later at the beginning of the first COS G160M sequence, EK~Dra apparently was in outburst, experiencing a giant FUV flare.  The star had brightened up in \ion{C}{4} by an order of magnitude, nearly triggering the COS detector bright limits, a remarkable performance for a nearly 8th-magnitude G dwarf.

Figure~4a illustrates time-resolved line fluxes and profiles from the first COS observation (visit 2), with grating G160M.  The top panel displays the total spectral image (all four FP-POS event lists combined), with specific integration bandpasses marked for the \ion{Si}{4} doublet (blue), \ion{C}{4} (red), and \ion{He}{2} (green); as well as two line-free continuum bands (``B:'' orange dashed; ``A:'' orange/black dashed); and two pairs of flanking background swaths (black dashed, and black-yellow dashed), one for each detector segment.  The parameters defining these measurements, and the subsequently described spectral profile extractions, are listed in Table~2.

The middle panel of Fig.~4a compares the time-resolved line and continuum count rates, with the same color-coding.  The individual points represent counts accumulated in sliding time bins to achieve a specified signal-to-noise ratio (S/N)$_{\rm crit}$: see Table~2.  The two continuum bands (orange, and black outlined orange, circles) were scaled down a factor of 4 for display purposes, while the background count rates (black and yellow dots; taken from each full FP-POS time interval) were scaled upward by 100 times relative to the continuum bands (taking into account the continuum integration wavelengths and cross-dispersion width: see Table~2) to illustrate the possible influence of the background for the worst-case scenario (broad wavelength bands).  The background corrections for the continuum bands in this instance are negligible, but still were compensated.  The background contributions for the tiny line spectral footprints are entirely negligible, and were not corrected.

The lower panels illustrate spectrally resolved profiles of three bright features -- \ion{Si}{4} $\lambda$1402,  \ion{He}{2} $\lambda$1640, and \ion{C}{4} $\lambda$1548 -- each accumulated in time intervals corresponding to a fixed number of counts ($N_{\rm crit}$: see Table~2).  \ion{Si}{4} $\lambda$1402 was chosen over stronger \ion{Si}{4} $\lambda$1393, because the latter is affected by bad pixels in one of the FP-POS segments of G160M/B.  The specific accumulation intervals are highlighted by small ticks in the middle panel time sequences.  In each of the bottom spectral panels, the time-resolved profiles are the thin colored curves.  The background yellow shaded envelope represents a profile of the same transition from a spectrum averaged over visits 3-5 (see below), scaled to the maximum and minimum of the collection of line peaks for the particular visit.  The upper and lower outlines of the envelopes are black dotted.  Remarkably, despite the large flux changes during the post-flare decline caught by visit~2, the profiles of all three features are nearly identical to the scaled averages; showing the characteristic redshift of the line core, and the high-velocity extended wings, especially on the redward side of line center.

Figure~4b is similar, for the G130M side A spectrum in the second part of visit~2.  Here, the integrated count rates and extracted profiles are for \ion{C}{2}, [\ion{Fe}{21}], and \ion{Si}{4}.  The two continuum bands (now smaller, to avoid the more numerous chromospheric emissions at these wavelengths) were combined into a single continuum flux, scaled down a factor of two in the figure for display.  The background count rates, again adjusted to the total $(\Delta{x},\,\Delta{y})$ continuum bandpass, were scaled upward a factor of 25 relative to the displayed continuum values.  All the fluxes continue to decline during this interval, but again the time-resolved line profiles are self-similar to the visits 3-5 averages.

Figure~5a illustrates the G160M sequence of visit 3, about 10 hours after the initial flare segment.  The continuum and background count rates are scaled as in Fig.~4a.  All the fluxes, especially the continuum bands, have fallen compared with visit 2, and now the background corrections are more significant.  The line profiles span a more compressed range, essentially identical to the visits 3-5 average.

Figure~5b depicts the G130M exposures of visit 3. The continuum and background fluxes are scaled as in Fig.~4b.  Again, there is little variability in the integrated fluxes or the time-resolved profiles.  The analogous diagrams for later visits 4 and 5 are very similar to those of visit 3, and are not illustrated.

\clearpage
\begin{figure}
\figurenum{4a}
\vskip  0mm
\hskip  7mm
\includegraphics[scale=0.75,angle=90]{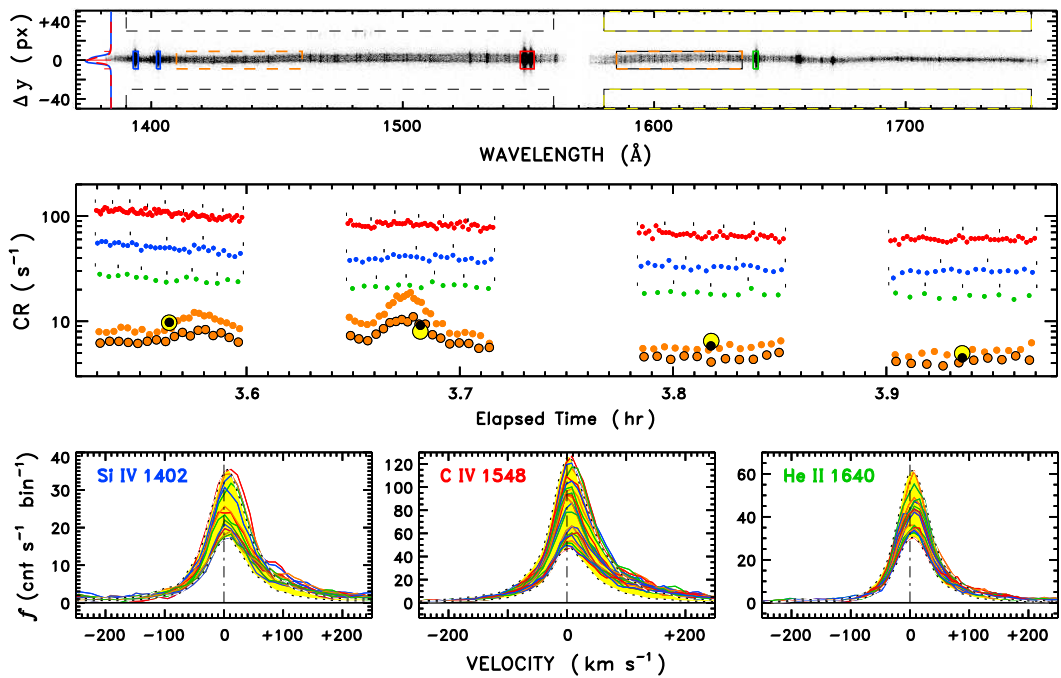} 
\vskip 0mm
\figcaption[]{\small Time-resolved line fluxes and profiles from the first COS observation (visit~2), during the large FUV flare, with grating G160M (see text for detailed description).
}
\end{figure}

\begin{figure}
\figurenum{4b}
\vskip  0mm
\hskip  7mm
\includegraphics[scale=0.75,angle=90]{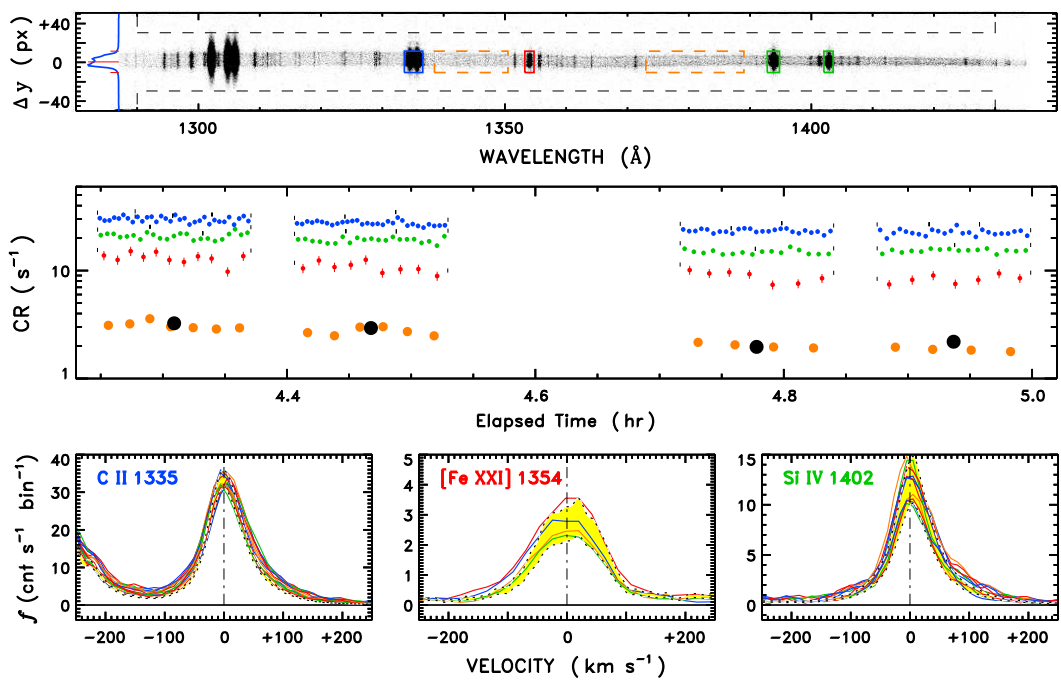} 
\vskip 0mm
\figcaption[]{\small Same as Fig.~4a, for the G130M side-A exposures during second part of visit~2.
}
\end{figure}

\clearpage
\begin{figure}
\figurenum{5a}
\vskip  0mm
\hskip  7mm
\includegraphics[scale=0.75,angle=90]{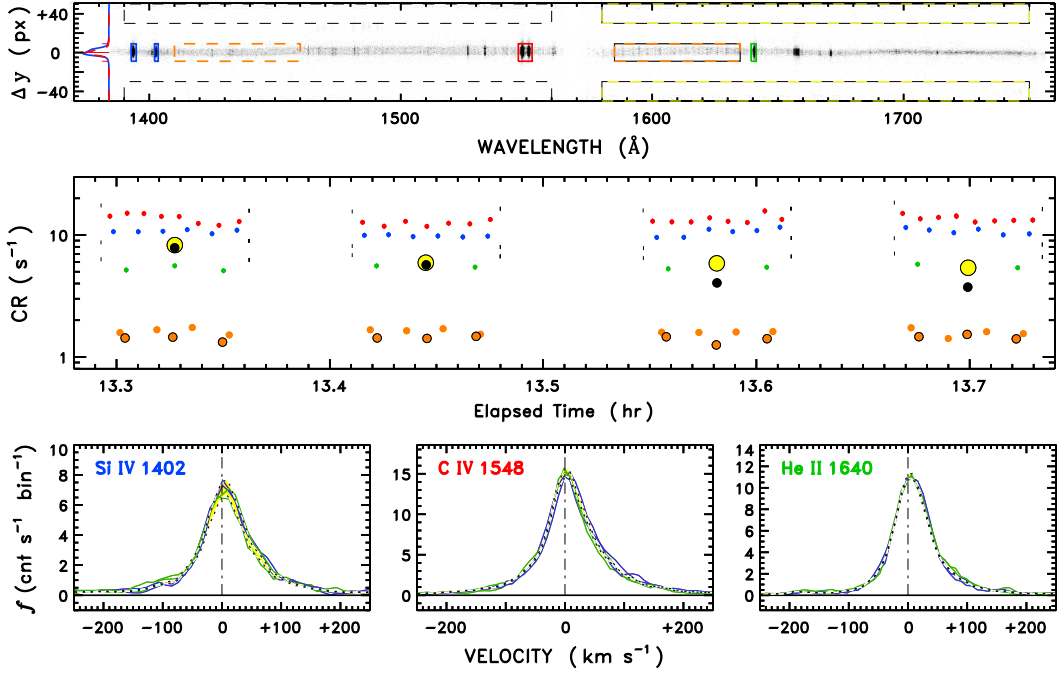} 
\vskip 0mm
\figcaption[]{\small Same as Fig.~4a for the G160M sequence of visit 3, about 10 hours after the initial flare segments.  Unlike Fig.~4a, there is little variability of the integrated fluxes or the time-resolved profiles.
}
\end{figure}

\begin{figure}
\figurenum{5b}
\vskip  0mm
\hskip  7mm
\includegraphics[scale=0.75,angle=90]{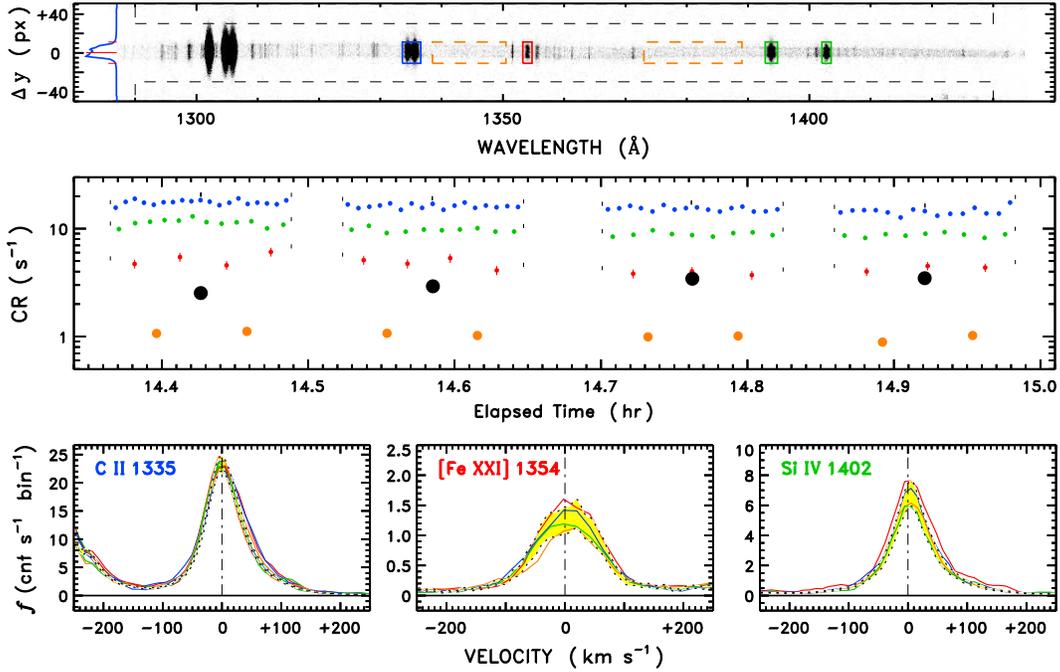} 
\vskip 0mm
\figcaption[]{\small Same as Fig.~4b for the G130M sequence of visit 3.  
}
\end{figure}

Figure~6 is an expanded view of the flare in \ion{Si}{4}, which was recorded in both COS grating settings.  As in Figs.~4 and 5, the symbols represent sliding time bins corresponding to (S/N)$_{\rm crit}= 20$.  Accordingly, the bins become wider in time, and the points more separated, as the count rates decline during the later stages of the event.  The COS \ion{Si}{4} values were normalized to the average of visits 3-5, which were much quieter than the visit~2 flare.  The G160M/B and G130M/A \ion{Si}{4} time series were separately normalized to the corresponding quiescent averages, to avoid possible differences in sensitivity with the different grating settings on the different detector segments.  The STIS \ion{Si}{4} points from visit~1, immediately prior to the flare rise, were based on calibrated fluxes integrated over the doublet components in each of the four independent E140M sub-exposures, divided by the integrated flux, measured in the same way, from an average COS spectrum constructed from the quiescent visits 3-5 (as detailed later).  The flare evolution was modeled by twin exponentials: a rapid initial decline with an e-folding time constant of 7~minutes, on top of a slower decay with a 1.9~hour time scale.  Such bimodal decays are common among stellar X-ray flares, for reasons that are not completely understood (see examples and discussion by Tsang et al.\ 2012).  The lower panel of Fig.~6 provides a broader view of the \ion{Si}{4} behavior during the entire {\em HST}\/ program, showing explicitly the relatively quiet intervals prior to and following the flare event.

\begin{figure}
\figurenum{6}
\vskip  0mm
\hskip -2mm
\includegraphics[scale=0.675,angle=90]{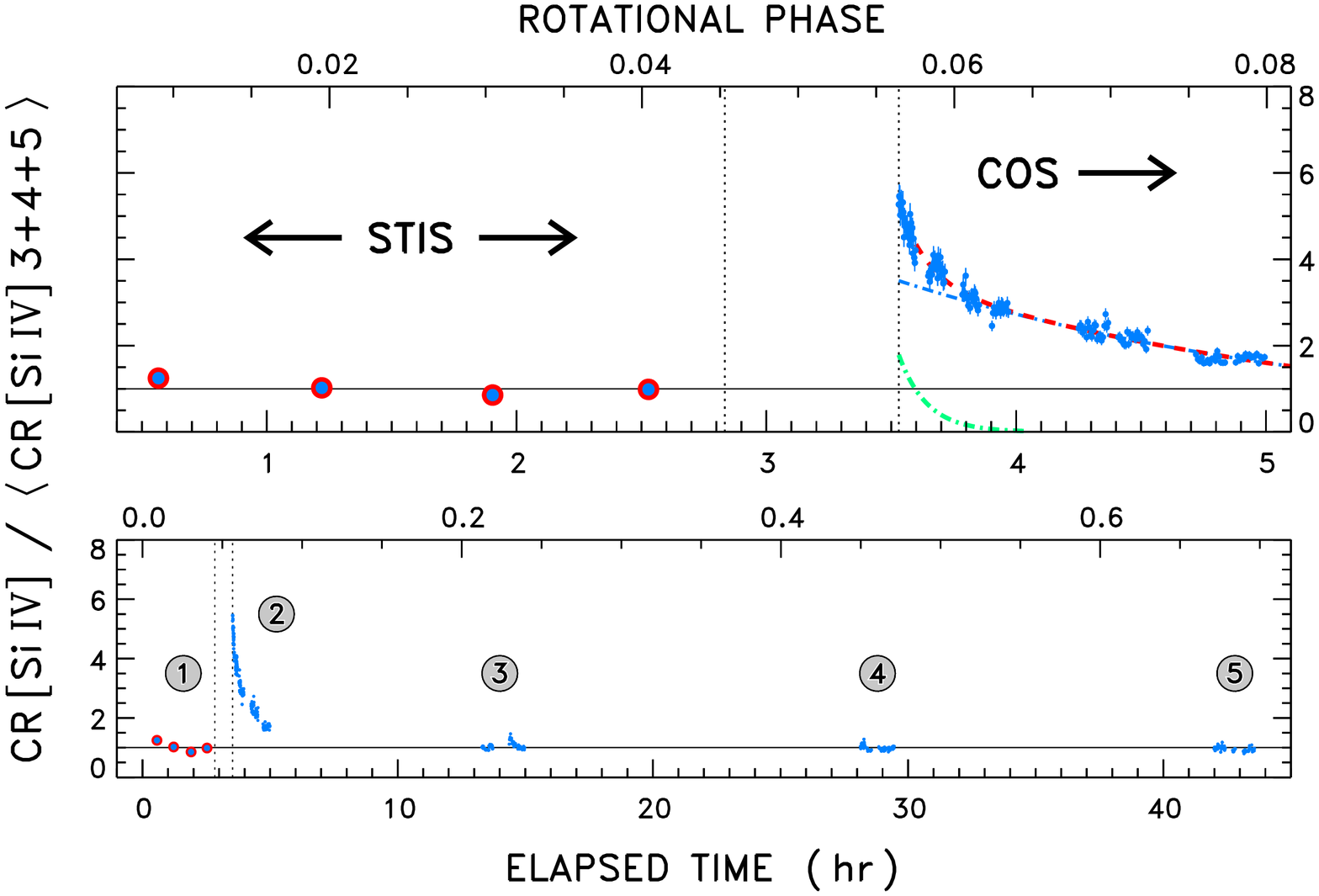} 
\vskip 0mm
\figcaption[]{\small \ion{Si}{4} $\lambda$1393\,+\,$\lambda$1402 time series covering visits 1 and 2 (top panel), and the full campaign (bottom panel).  Origin of the timeline is the initial STIS target acquisition.  The \ion{Si}{4} count rates were normalized to the average of visits 3-5.  Right side of the upper panel contains a close-up view of the serendipitous large FUV flare, captured in \ion{Si}{4} by the sequential grating settings (each split into four FP-POS sub-exposures).  The flare decay can be described as an initial rapid exponential drop on the scale of minutes (green dotted-dashed curve), followed by a slower exponential decline on the scale of hours (blue dotted-dashed; red dashed is sum).  Left side of the upper panel depicts the four STIS sub-exposures of visit~1.  Lower panel portrays the 1.5-hour event within the context of the pre-flare (STIS visit 1) and post-flare (COS visits 3-5) quiescent periods.
}
\end{figure}

Figure~7 is a multi-spectral view of the EK~Dra flare decay, and subsequent quiescent intervals. Error bars to the left of the vertical dotted line indicate the STIS average levels and standard deviations (over the four sub-exposures). The COS time series to the right are the same as illustrated in Figs.~4 and 5, but now including visits 4 and 5, and normalized to the visits 3-5 average.  Note the small flare decay that affects all the features in the second part of visit~3, together with a number of \ion{Si}{4} and \ion{C}{4} transients like those seen in the Saar \& Bookbinder time series and the 2010 SNAP program.  Also note the two bursts (at $t= $3.6~hr and 3.7~hr) in the G160M blue and red continuum bands (marked by the corresponding colors), which remarkably are not repeated in any of the other activity tracers.  This phenomenon is discussed in more detail later (\S{3.4}).

\begin{figure}
\figurenum{7}
\vskip  0mm
\hskip  -15mm
\includegraphics[scale=0.75,angle=90]{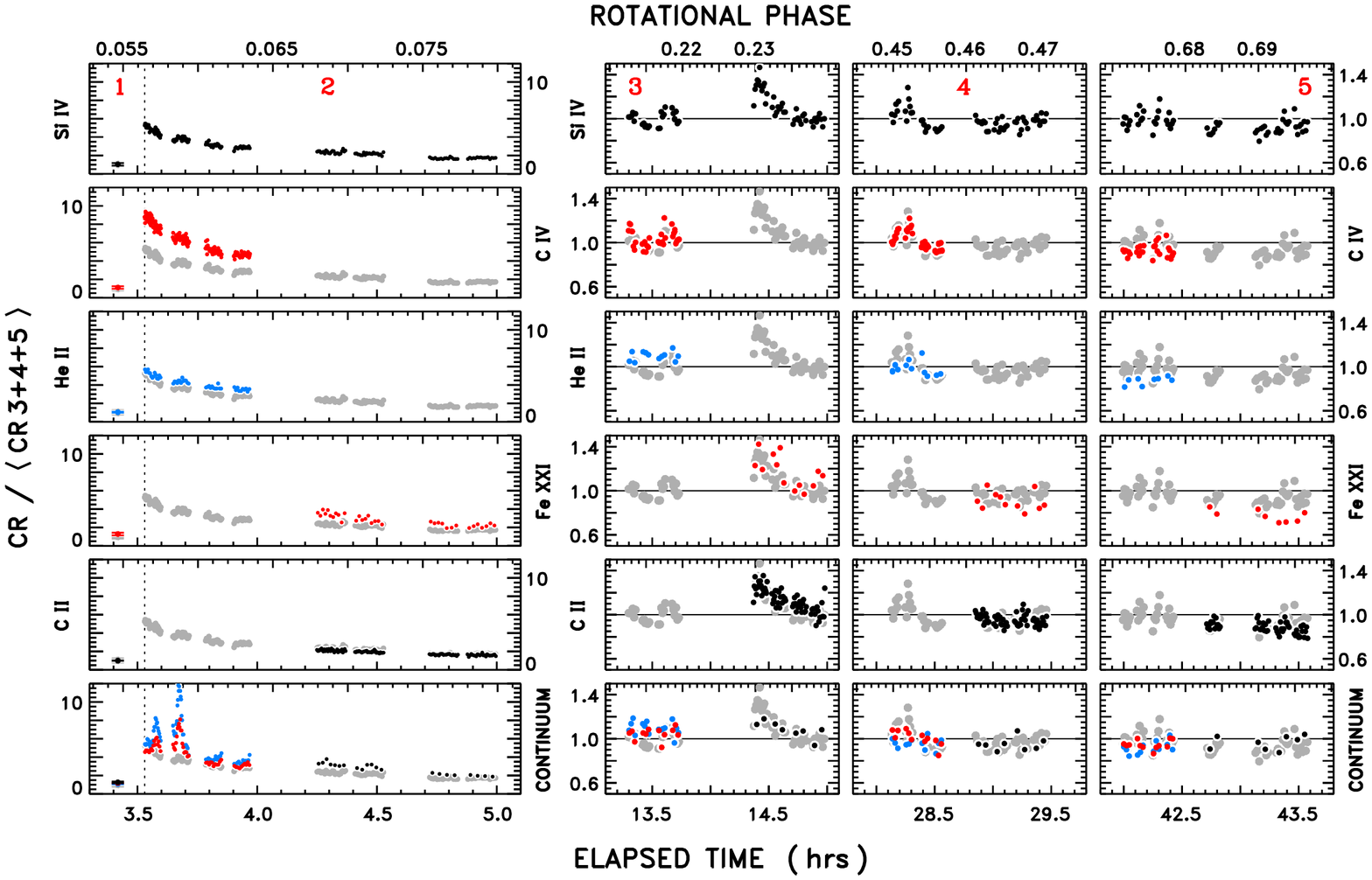} 
\vskip 0mm
\figcaption[]{\small Multi-spectral view of the EK~Dra flare decay, and subsequent quiescent intervals. Error bars to left of the dotted line (which marks beginning of visit 2) indicate the STIS average levels.  Visit numbers are in red in the upper panels.  Count rates of each species were normalized to the visits 3-5 average.  Larger gray symbols in each panel (except top-most) are the \ion{Si}{4} time series repeated for comparison.
}
\end{figure}

Although the large FUV flare seemingly ruined the carefully orchestrated STIS-COS velocity cross-calibration, the bright side is that this was the first such significant event on a sunlike star to have been captured with the un-matched spectroscopic and time-resolution capabilities of an instrument like COS; a unique opportunity to explore properties of the outburst beyond the usual simple intensity evolution (with modest temperature discrimination) delivered by, say, an X-ray telescope.  And, as described above, following the large flare EK Dra settled into an period of apparent quiescence -- to the extent that a star with a thousand times the X-ray luminosity of the Sun can be considered ``quiet'' -- for the next three COS visits (spanning a day and a half, corresponding to about half a rotation period).  The similarity of the time-averaged FUV spectra of these three visits, both among themselves and compared with the initial STIS pointing, ultimately provided the desired velocity cross-validation (as described later in \S{3.1}).  So, all was not lost.

Furthermore, the lack of significant FUV profile variability of the hot lines between STIS visit~1 and COS visits~3-5, implies that ``Doppler Imaging'' was not an important factor during the 2012 {\em HST}\/ campaign.  In fact, the constancy of the hot-line fluxes over the more than half a rotational cycle also indicates that the surface distribution of FUV-emitting regions was relatively uniform, at least in that epoch.  At higher energies, in contrast, G\"udel et al.\ (1995) reported factor of $\sim$2 systematic variations in {\em ROSAT}\/ all-sky survey X-ray fluxes, in step with the rotation period, and a suggestion of correlated variability at microwave frequencies as well.  On the Sun, albeit perhaps not the best guide to hyperactive EK~Dra, the FUV-bright plage regions surrounding sunspot groups are much more spatially pervasive than the dark spots themselves, and more so than the compact coronal magnetic loop systems associated with the active region as well.

\subsubsection{EK~Draconis B}

EK~Dra is reported to have a faint, M-type companion in a highly eccentric 45~year orbit (K{\"o}nig et al.\ 2005).  Based on the authors' ephemeris, the secondary would have been just past maximum elongation in 2012, nearly due south of the primary with a separation of $0.75^{\prime\prime}$.  This object would be at the inner edge of the COS PSA, and potentially visible as a slightly displaced spectrum (unfortunately on top of the main spectrum in the $+x$ direction, given the position angle of the COS focal plane at the time of the 2012 program), but entirely outside the STIS $0.2^{\prime\prime}{\times}0.2^{\prime\prime}$ slot.  However, there is no obvious line doubling due to a second spectrum visible in any of the COS FUV grating images; although the predicted separation between the two is only +0.35~\AA\ (EK~Dra B to red of A), and the bright primary lines are broad.

The secondary also was not clearly visible in (stacked versions) of either the STIS or COS acquisition images of the 2012 campaign.  Serendipitously, however, the faint companion was recovered in a subsequent STIS pointing on EK~Dra, in an 2014 ISM SNAP program (S.\ Redfield P.I.; see Table~1).  The partial orbit visit had a more conservative CCD acquisition: a total exposure time of 2.3~s through the F25ND3 aperture, significantly deeper than the 0.3~s of the 2012 program with the same filter.  The secondary was about 1\% of the primary's brightness, at a position angle of 172$\degr$ and separation of 0.75$^{\prime\prime}$, just as predicted by the ephemeris of K{\"o}nig et al.\ (2005).  The $V$ magnitude of the secondary would be about +13.5, taking into account that the broad-band CCD response favors red stars over yellow stars.  The primary/secondary $\Delta{V}\sim 6$ is what was suggested by K{\"o}nig et al.
  
This is important, because a youthful red dwarf could be quite active, and its hot lines, if bright enough, potentially could affect the profiles of the primary features (with the unfortunate location of B on top of the A spectrum in 2012).  However, scaling FUV fluxes of the most active dMe flare stars, like AU~Microscopii and AD Leonis, to the estimated $V$ magnitude of the secondary predicts peak flux densities, say at \ion{C}{4}, of only a few percent of the primary's, which would be of little concern.  A negligible impact by the red star, at least on the quiescent FUV spectrum of EK~Dra, is supported by the fact that the COS visits~3-5 spectra are so similar to STIS visit~1, which entirely excluded the secondary by virtue of the small STIS aperture.

As for the large FUV flare, if it were from the red star, the associated soft X-ray luminosity (see below) would have been quasi-bolometric (i.e., $L_{\rm X}/L_{\rm bol}\sim 1$); not impossible, but very unlikely.  The $\sim$2 known examples, such as the quasi-bolometric flare on DG~CVn reported by Drake et al.\ (2014), have come from the {\em Swift}\/ Burst Alert Telescope, and soft-X-ray follow-up by that satellite.  The survey burst alert situation is much more likely to capture extremely rare events because it is viewing a whole sky full of objects, all the time, rather than pointed at a single target for a limited period.

Furthermore, the enhanced narrow peaks of the FUV hot lines during the flare were nearly identical in shape and velocity to the quiescent profiles of the G star seen outside the outburst; rather than shifted by the $\sim$+75 km s$^{-1}$ to the red expected if the FUV lines were instead from the secondary star.  The preponderance of evidence is that the large FUV flare was from the G-type primary.

\section{ANALYSIS}

\subsection{STIS-COS Wavelength Cross-Calibration}

A main objective of the 2012 EK~Dra campaign was to test the reality of the large $\sim 20$~km s$^{-1}$ Doppler shifts of \ion{Si}{4} recorded in the brief COS SNAPshot two years earlier.  The core of that test was to validate the COS wavelengths against the highly precise scales of STIS.  In particular, the E140M-1425 echelle setting captures the entire FUV region (1150-1700~\AA) in a single shot, so there is no issue trying to tie together multiple spectral segments (e.g., G130M and G160M of COS, including the individual FP-POS steps, each with its own potential calibration issues).  Further, the STIS MAMA cameras have hard-wired physical pixels, whereas the cross-delay-line detectors of COS have floating pixels defined by the $(x,y)$ impact point of the electron cloud associated with a photon event.  The implication for STIS is that a given pixel always experiences whatever geometrical distortions are present locally, and these can be accounted to some degree in the wavelength dispersion solution.  With COS, conversely, the geometrical distortions are sampled quasi-randomly by whatever distribution of impact points happen to fall at a particular location.

Perhaps more important, the STIS wavelength calibration light (from onboard hollow-cathode emission lamps) follows the same optical path as an external stellar point source, falling on the same echelle order pixels as the stellar spectrum, and consequently experiences the same geometrical distortions.  Thus, the STIS wavelength calibration spectra encode significant information concerning these local geometrical distortions, which then can be accommodated in a hybrid dispersion solution, such as described in the ASTRAL protocols mentioned earlier.  For COS, on the other hand, the lamp-flash spectrum falls well above the spectral stripe (see, e.g., Fig.~3), on a portion of the detector where the geometrical distortions are completely different.  Consequently, the COS lamp-flashes are less relevant for confirming the pipeline dispersion solutions.

To test the wavelength fidelity of COS, the three independent combinations of grating setting and detector segment were considered separately.  Reference spectra were taken from the {\large\sf calcos} pipeline ``x1d'' files, which include a summation over the four FP-POS exposures, eliminating bad pixels and suppressing fixed pattern noise such as grid-wire shadows.  The visits 3-5 spectra for each grating setting were registered to visit~3 by cross-correlation, and co-added to increase S/N.  In all cases, extracted lamp-flash lines from the individual FP-POS exposures were very stable in velocity, demonstrating the repeatability of the {\large\sf calcos} pipeline wavelength assignments (although not necessarily their accuracy).

Figure~8 compares selected profiles of bright lines from the various COS grating/detector-segment combinations against the STIS E140M spectrum from visit~1, convolved with the COS line-spread function.  The zero point of the STIS wavelength scale (as in Fig.~2) was derived from the average velocity of \ion{O}{1} $\lambda$1304, $\lambda$1306, and \ion{C{\em l}}{1} $\lambda$1351, the brightest low-excitation narrow chromospheric lines recorded by STIS.  The ground-state \ion{O}{1} $\lambda$1302 resonance line was avoided owing to a blue asymmetry imposed by redward interstellar atomic oxygen absorption (see Fig.~9).  Note also that the STIS \ion{A{\em l}}{1} $\lambda$1670 resonance line is truncated in its red wing by a gap between adjacent echelle orders.  The derived chromospheric radial velocity was $-23$~km s$^{-1}$, compared with the photospheric value of $-20$~km s$^{-1}$ expected in 2012 near orbital phase 0.5 (K\"onig et al.\ 2005).  This indicates that part of the velocity correction, amounting to $-3$~km s$^{-1}$, could be a zero-point error in the STIS wavelength scale for this observation, assuming that the FUV chromospheric velocity measured by STIS, and the optical photospheric velocities measured by K{\"o}nig et al., refer to the same frame, and further that the optical measurements are freer of systematics.  Aside from the possible zero-point correction, the STIS wavelength scales have high internal precision.

\begin{figure}
\figurenum{8}
\vskip  -11mm
\hskip   +2.5mm
\includegraphics[scale=0.800]{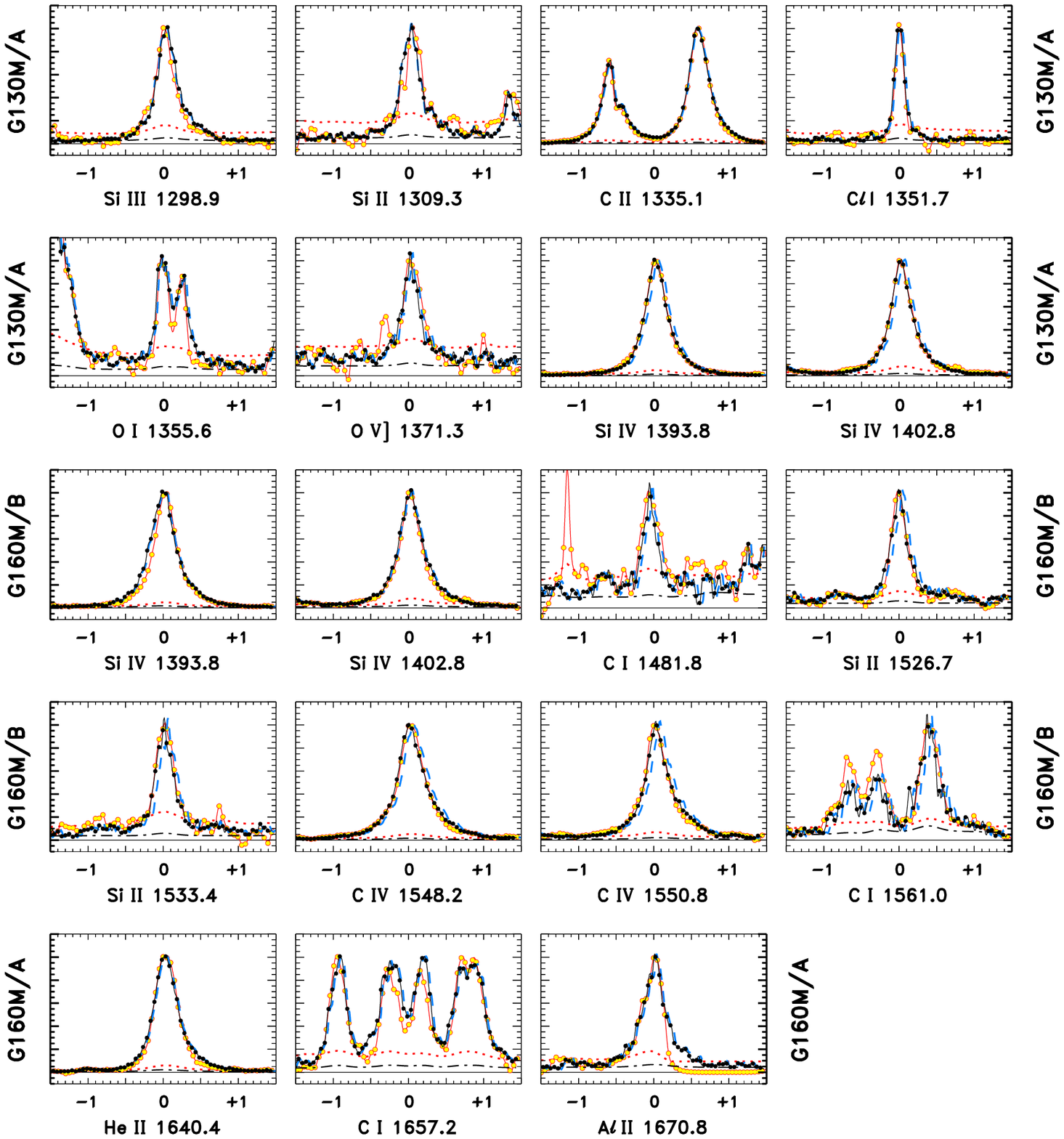} 
\vskip -8mm
\figcaption[]{\small Selected intervals from the various COS grating/detector-segment combinations, averaged over quiescent visits 3-5, compared to the STIS visit 1 spectrum, smoothed to COS resolution.  The $x$-axis scale in each panel is $\Delta\lambda$ (\AA) relative to the wavelength cited in the axis title.  The STIS spectra, in the stellar frame defined by narrow chromospheric emission lines, are represented by red curves and open circles; the red dotted curves are the smoothed 1\,$\sigma$ photometric noise (per resel).  Blue dashed traces are for the uncorrected COS pipeline wavelengths, but accounting for the predicted photospheric radial velocity.  Black dotted curves are for a simple modification of the COS wavelength scales to force better global agreement with STIS.  Black dotted-dashed curves are smoothed COS 1\,$\sigma$ photometric errors.  For the STIS and COS spectra, only every fifth point is plotted.  The narrow spike in the \ion{C}{1} $\lambda$1481 panel is a STIS hot pixel.}
\end{figure}

The COS spectra are depicted in two ways: (1) with the original pipeline wavelengths, adjusted only for the assumed stellar radial velocity ($-20$~km s$^{-1}$); and (2) for a simple modification of the COS wavelength scales to force better global agreement with STIS.  The latter corrections, expressed as equivalent velocities, are of the form:
\begin{equation}
\Delta{\upsilon}= \upsilon_{0} + a\, (\lambda - \lambda_{0})\,\,\,   ,
\end{equation} 
The value $\upsilon_{0}= -3$~km s$^{-1}$ was used for all the grating/segment combinations.  The $a$ coefficients and reference wavelengths, $\lambda_{0}$, are as follows: $-$0.112~km s$^{-1}$ \AA$^{-1}$ and 1340~\AA\ for G130M/A; $-$0.054~km s$^{-1}$ \AA$^{-1}$ and 1400~\AA\ for G160M/B; and $-$0.046~km s$^{-1}$ \AA$^{-1}$ and 1640~\AA\ for G160M/A.  Here, the negative $a$ coefficients are compensating for systematic redshifts of the assigned wavelengths that become progressively larger with wavelength.  For example, the \ion{Si}{4} $\lambda$1393 feature recorded by G130M/A (as in the earlier Ayres \& France COS SNAPshot), would have a total instrumental redshift of 10~km s$^{-1}$, about half of the \ion{Si}{4} velocities ($\sim 20$~km s$^{-1}$) measured in that study.  However, a direct comparison between the two is subject to unknown systematic errors, not the least due to the different acquisition techniques, so the difference should not be taken too literally.  

The corrected COS spectra, in general, agree well with ground-truth STIS, although there is at least one important exception -- \ion{Si}{4} $\lambda$1393 at the extreme blue edge of G160M/B -- where the linear velocity model appears to fall short.  The cited grating-dependent velocity corrections were applied to the time-resolved COS profiles illustrated earlier in Figs.~4 and 5.  Nevertheless, the fundamental velocity measurements of the brightest lines, at least for the quiescent periods of EK~Dra, were derived from the more precise STIS spectra, as described below.

\subsection{Interstellar Features}

As mentioned earlier, there are interstellar absorptions present in several of the ground-state resonance lines of the STIS FUV spectrum of EK~Dra.  These offer the opportunity to check the STIS zero-point velocity deduced from narrow chromospheric emissions.  The ISM features are redshifted with respect to EK~Dra, with an average heliocentric velocity of about $-$2~km s$^{-1}$.  According to the maps of Redfield \& Linsky (2008), EK~Dra falls at the edge of the Local Interstellar Cloud (LIC) and also at the edge of the smaller North Galactic Pole (NGP) cloud.  Depending on the velocity vector assumed for the LIC (e.g., Gry \& Jenkins 2014), the predicted heliocentric velocity in the direction of EK~Dra falls in the range $-$(1-3)~km s$^{-1}$; while for the NGP cloud, the projected velocity is close to 0~km s$^{-1}$: the galactic coordinates of EK~Dra are nearly perpendicular to both directions of motion.  It also is possible, and perhaps likely, that both clouds are contributing to the ISM absorptions in the EK~Dra spectrum.

Weighing in on the matter is a recent STIS NUV high-resolution echelle (E230H-2713) spectrum of EK~Dra, taken in the 2014 ISM SNAP program mentioned earlier: a 600~s exposure through the $0.2^{\prime\prime}{\times}0.2^{\prime\prime}$ slot, without a peak-up (see Table~1).  The spectrum covers the important \ion{Mg}{2} resonance lines at 2796~\AA\ and 2803~\AA, which have ISM absorptions on the red sides of their broad chromospheric emission cores.  The \ion{Mg}{2} ISM features fall at a velocity of $-2$ km s$^{-1}$ heliocentric.  However, cross-correlation against the same E230H-2713 bandpass from $\alpha$~Cen A suggests that there might be a $-1$~km s$^{-1}$ systematic error in the EK~Dra NUV photospheric spectrum, which would revise the ISM velocity to about $-1$~km s$^{-1}$.  The FUV and NUV interstellar features are illustrated in Figure~9.  The agreement between the two independent sets of ISM velocities provides additional confidence in the zero-point velocity calibration of the STIS FUV spectra.

\begin{figure}
\figurenum{9}
\vskip  0mm
\hskip  24mm
\includegraphics[scale=0.625]{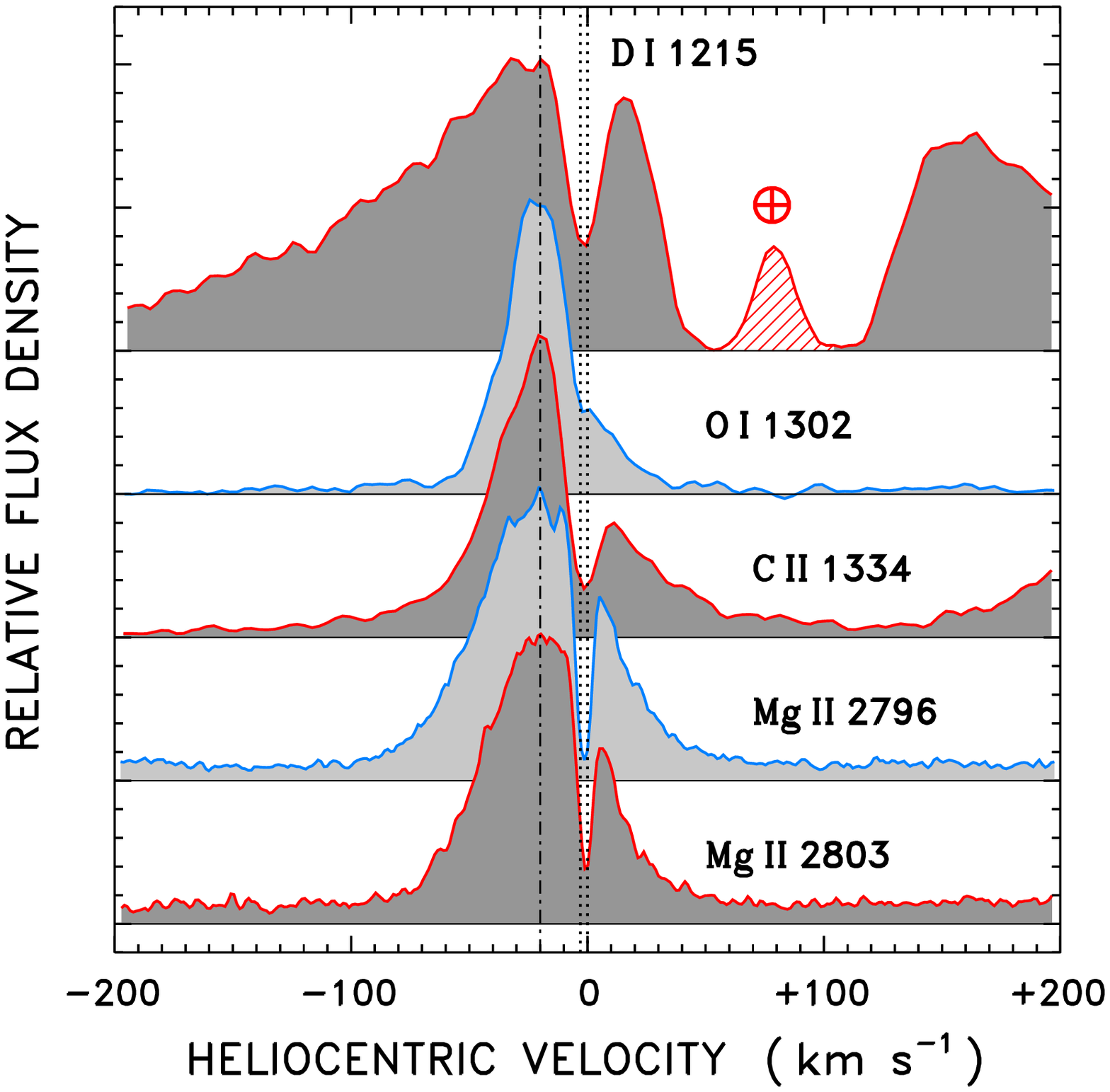} 
\vskip 0mm
\figcaption[]{\small Montage of ISM absorptions in the cores of strong chromospheric emissions of EK~Dra, from the STIS FUV and NUV spectra.  Each profile was normalized to the local flux density maximum in the velocity range, and separated vertically for clarity.  The hatched emission feature in the center of the Ly$\alpha$ absorption core is \ion{H}{1} skyglow.  The dotted-dashed line marks the predicted stellar radial velocity (not relevant for \ion{D}{1} because \ion{H}{1} dominates the emission profile), which was the same in the two separate epochs (FUV in 2012, NUV in 2014).  Dotted lines delimit the range of predicted heliocentric velocities for the LISM and NGP clouds in the direction of EK~Dra, which happens to fall at the edges of both structures, and nearly orthogonal to the respective velocity vectors.  The NUV \ion{Mg}{2} components are narrower than the FUV counterparts because the STIS high-resolution echelle was used at the longer wavelengths, and the thermal broadening is less for the heavier ion.
}
\end{figure}

\subsection{STIS and COS Spectral Measurements}

Tables~3-5 list representative measurements from STIS visit~1 (Table~3), the COS visits 3-5 averages for both G130M and G160M (Table~4), and the COS visit~2 flare (Table~5: G160M only).  The STIS entries include important bright lines below the shortwavelength cutoff of COS G130M/A at 1290~\AA, such as \ion{H}{1} Ly$\alpha$, \ion{Si}{3} $\lambda$1206, and the \ion{N}{5} resonance doublet at 1240~\AA.   STIS also recorded detailed profiles of the \ion{O}{1} 1305~\AA\ triplet lines, which fall at the blue end of G130M/A but are corrupted by atomic oxygen skyglow in most of the COS pointings.  The COS Tables include weaker features that were not detectable in the STIS spectrum, especially the key density-diagnostic multiplet of \ion{O}{4}] in the 1400~\AA\ region.  The same G160M features in Table~4 also were measured in the integrated G160M flare spectrum of visit~2, and reported in Table~5, to compare with the more quiescent intervals following the outburst.

For COS, the measured Doppler shifts are affected by residual errors in the simple correction of the G130M and G160M wavelength scales, so should be treated with caution.  However, the absolute fluxes and derived line widths, say of narrow components (NC) and broad components (BC) in a bimodal fit, should be more reliable, since neither depends sensitively on accurate wavelengths.  Any systematic Doppler shifts between NC and BC within a given spectral feature should be reliable as well, because the (corrected) COS wavelengths have sufficient precision over small wavelength separations.

\clearpage
\begin{deluxetable}{llccccc}
\tablenum{3}
\tabletypesize{\footnotesize}
\tablecaption{{\em HST}\/ Measurements: STIS FUV and NUV}
\tablecolumns{7}
\tablewidth{0pt} 
\tablehead{
\colhead{Species} & \colhead{$\lambda_{\rm lab}$} & \colhead{Flag} & \colhead{$\upsilon_{\rm L}$} & \colhead{FWHM$_{\rm L}$} & 
\colhead{$f_{\rm L}$} & \colhead{$f_{\rm C}$}\\
\colhead{} & \colhead{(\AA)} & \colhead{} & \multicolumn{2}{c}{(km s$^{-1}$)} &  
\colhead{(10$^{-14}$ cgs)} &
\colhead{(10$^{-14}$ cgs \AA$^{-1}$)}\\
\colhead{(1)} & \colhead{(2)} & \colhead{(3)} & \colhead{(4)} & \colhead{(5)} & \colhead{(6)} & 
\colhead{(7)}
}
\startdata
\ion{C}{3} & 1175{\em m} &   INT & (1175.7~\AA) & (2.2~\AA) & $     6.1$ & 
    0.00 \\
\ion{Si}{3} & 1206.500 &  GAU2 & $+6{\pm}   1$ & $   69{\pm}   5$ & 
$     4.3{\pm}   0.5$ &     0.00 \\
\ion{Si}{3} & 1206.500 &  GAU2 & $+17{\pm}   8$ & $  184{\pm}  33$ & 
$     2.8{\pm}   0.5$ &  \nodata  \\
\ion{D}{1} & 1215.339 & GAU1S & $+18{\pm}   0$ & $   13{\pm}   1$ & 
$    -3.3{\pm}   0.3$ & $   67,  -44$  \\
\ion{H}{1} & 1215.670 &  GAU2\tablenotemark{a} & $ -8{\pm}   1$ & $  (231{\pm}   7)$ & 
$  (71{\pm}   2)$ &     0.43 \\
\ion{H}{1} & 1215.670 &  GAU2\tablenotemark{a} & $ -4{\pm}   4$ & $  (565{\pm}  21)$ & 
$  (32{\pm}   2)$ &  \nodata  \\
\ion{H}{1} & 1215.670 &  GAU1\tablenotemark{a} & $ -9.3{\pm}  0.7$ & $  (272{\pm}   3)$ & 
$  (94{\pm}   1)$ &     0.29 \\
\ion{H}{1} & 1215.670 &  GAU1X & $+20{\pm}   0$ & \nodata & 
\nodata &  \nodata \\
\ion{H}{1} & 1215.670 &   INT & (1215.6~\AA) & (5.0~\AA) & $  67$ &     0.43 \\
\ion{N}{5} & 1238.810 & GAU2D & $+4{\pm}   1$ & $   47{\pm}   5$ & 
$    0.81{\pm}  0.12$ &     0.00 \\
\ion{N}{5} & 1238.810 & GAU2D & $+8{\pm}   6$ & $  157{\pm}  25$ & 
$    0.97{\pm}  0.11$ & [$  0.477,     -2.4$] \\
\ion{O}{1} & 1302.168 & GAU2R & $+3.9{\pm}  0.7$ & $   36{\pm}   1$ & 
$     1.6{\pm}   0.1$ &     0.00 \\
\ion{O}{1} & 1302.168 & GAU2R & $+16{\pm}   1$ & $   15{\pm}   1$ & 
$   -0.35{\pm}  0.12$ &  \nodata  \\
\ion{O}{1} & 1304.858 &  GAU1 & $ -0.5{\pm}  0.5$ & $   33{\pm}   1$ & 
$     1.2{\pm}   0.0$ &     0.00 \\
\ion{O}{1} & 1306.029 &  GAU1 & $+0.0{\pm}  0.5$ & $   31{\pm}   1$ & 
$     1.2{\pm}   0.0$ &     0.00 \\
\ion{C}{2} & 1334.532 & GAU2R & $+6.1{\pm}  0.4$ & $   51{\pm}   1$ & 
$     4.5{\pm}   0.1$ &     0.02 \\
\ion{C}{2} & 1334.532 & GAU2R & $+17{\pm}   0$ & $   14{\pm}   1$ & 
$    -1.2{\pm}   0.2$ &  \nodata  \\
\ion{C}{2} & 1335.702 &  GAU2 & $+3.0{\pm}  0.4$ & $   48{\pm}   2$ & 
$     3.7{\pm}   0.2$ &     0.02 \\
\ion{C}{2} & 1335.702 &  GAU2 & $+11{\pm}   2$ & $  137{\pm}   8$ & 
$     2.9{\pm}   0.2$ &  \nodata  \\
\ion{C}{2} & 1334{\em m} &   INT & (1335.2~\AA) & (2.8~\AA) & $  11$ &     0.02
 \\
\ion{C{\em l}}{1}$\ast$ & 1351.656 &  GAU1 & $+0.4{\pm}  0.9$ & $   20{\pm}   2$
 & $    0.23{\pm}  0.02$ &     0.03 \\
\ion{Fe}{21} &  1354.08 &  GAU1 & $ -4{\pm}   3$ & $  103{\pm}   8$ & 
$    0.68{\pm}  0.04$ &     0.03 \\
\ion{O}{1} & 1355.598 &  GAU1 & $ -1{\pm}   2$ & $   27{\pm}   4$ & 
$    0.13{\pm}  0.01$ &     0.03 \\
\ion{C}{1} & 1355.844 &  GAU1 & $ 0{\pm}   2$ & $   27{\pm}   4$ & 
$    0.10{\pm}  0.01$ &     0.00 \\
\ion{Si}{4} & 1393.760 & GAU2D & $+6.2{\pm}  0.6$ & $   51{\pm}   2$ & 
$     2.7{\pm}   0.1$ &     0.02 \\
\ion{Si}{4} & 1393.760 & GAU2D & $+10{\pm}   3$ & $  174{\pm}   8$ & 
$     2.6{\pm}   0.1$ & [$  0.546,     -0.9$] \\
\ion{C}{4} & 1548.204 & GAU2D & $+7.1{\pm}  0.6$ & $   50{\pm}   2$ & 
$     5.6{\pm}   0.3$ &     0.03 \\
\ion{C}{4} & 1548.204 & GAU2D & $+11{\pm}   2$ & $  179{\pm}   8$ & 
$     6.9{\pm}   0.3$ & [$  0.585,     -2.3$] \\
\ion{He}{2} & 1640.404 &  GAU2 & $+5.8{\pm}  0.7$ & $   54{\pm}   3$ & 
$     8.3{\pm}   0.7$ &     0.11 \\
\ion{He}{2} & 1640.404 &  GAU2 & $+28{\pm}  14$ & $  132{\pm}  36$ & 
$     2.4{\pm}   0.7$ &  \nodata  \\
\cutinhead{STIS Visit 1: \ion{N}{5}\,+\,\ion{Si}{4}\,+\,\ion{C}{4}}
\multicolumn{2}{l}{Narrow component}  &   GAU2 &  $  +5.1{\pm} 0.1$  & $    50{\pm}    1$ & $  0.48{\pm} 0.01$ &  \nodata \\
\multicolumn{2}{l}{Broad component}  &   GAU2 &  $  +8.7{\pm} 0.4$  & $   137{\pm}    1$ & $  0.52{\pm} 0.01$ &  \nodata \\
\cutinhead{\ion{Mg}{2} Chromospheric Emission Cores in Stellar Rest Frame}
\ion{Mg}{2} & 2796.352 &  GAU1 & $ +0.1{\pm}  0.3$ & $   60.0{\pm}  0.9$ & 
$    150{\pm}  2$ &     13.2 \\
\ion{Mg}{2} & 2803.531 &  GAU1 & $ +0.3{\pm}  0.3$ & $   54.3{\pm}  0.6$ & 
$    117{\pm}  2$ &     13.2 \\
\cutinhead{\ion{Mg}{2} Interstellar Lines in Heliocentric Frame}
\ion{Mg}{2} & 2796.352 &  GAU1S & $ -1.8{\pm}  0.1$ & $   5.9{\pm}   0.3$ & 
$    -13.8{\pm}  0.6$ &    $201, -531$ \\
\ion{Mg}{2} & 2803.531 &  GAU1S & $ -1.2{\pm}  0.1$ & $   5.4{\pm}   0.3$ & 
$    -8.6{\pm}  0.4$ &     $166, -499$
\enddata
\tablecomments{Cols.~1 and 2 list the species and laboratory (vacuum) wavelength: suffix ``{\em m}'' indicates a multiplet.  Col.~3 is a processing flag: see text for specific descriptions.  Cols.~4 and 5 are centroid and width of the Gaussian profile, in velocity units.  Col.~6 is the integrated flux, in units of 10$^{-14}$ erg cm$^{-2}$ s$^{-1}$.  Col.~7 usually is the derived continuum flux density, with units of 10$^{-14}$ erg cm$^{-2}$ s$^{-1}$ \AA$^{-1}$.  If two values are listed, the second value is the gradient of a sloping fit (see text).  Note, in the specific case of a ``GAU2D'' constrained doublet fit, the first set of entries are the continuum parameters, but the second, bracketed, set is the ratio of the integrated flux of the secondary member divided by the principal member, and the deviation of the measured wavelength difference relative to the laboratory value (in km s$^{-1}$).  An error of zero indicates an uncertainty less than the least significant figure of the measurement.} 
\tablenotetext{a}{Fit ignores deep ISM central reversal, solely to obtain centroid velocity of outer profile.}
\end{deluxetable}

\clearpage
\begin{deluxetable}{llccccc}
\tabletypesize{\footnotesize}
\tablenum{4} 
\tablecaption{{\em HST}\/ Measurements: COS Visits 3-5 (Quiescent)} 
\tablecolumns{7}
\tablewidth{0pt} 
\tablehead{
\colhead{Species} & \colhead{$\lambda_{\rm lab}$} & \colhead{Flag} & \colhead{$\upsilon_{\rm L}$} & \colhead{FWHM$_{\rm L}$} & 
\colhead{$f_{\rm L}$} & \colhead{$f_{\rm C}$}\\
\colhead{} & \colhead{(\AA)} & \colhead{} & \multicolumn{2}{c}{(km s$^{-1}$)} &  
\colhead{(10$^{-14}$ cgs)} &
\colhead{(10$^{-14}$ cgs \AA$^{-1}$)}\\
\colhead{(1)} & \colhead{(2)} & \colhead{(3)} & \colhead{(4)} & \colhead{(5)} & \colhead{(6)} & 
\colhead{(7)} 
}
\startdata
\cutinhead{G130M}
\ion{C}{2} & 1334.532 & GAU2R & $+9.7{\pm}  0.0$ & $   60{\pm}   0$ & 
$     4.5{\pm}   0.0$ &     0.03 \\
\ion{C}{2} & 1334.532 & GAU2R & $+22{\pm}   0$ & $   17{\pm}   0$ & 
$   -0.97{\pm}  0.04$ &  \nodata  \\
\ion{C}{2} & 1335.702 &  GAU2 & $+0.2{\pm}  0.2$ & $   48{\pm}   0$ & 
$     3.2{\pm}   0.0$ &     0.03 \\
\ion{C}{2} & 1335.702 &  GAU2 & $+16{\pm}   0$ & $  122{\pm}   1$ & 
$     3.4{\pm}   0.1$ &  \nodata  \\
\ion{C}{2} & 1334{\em m} &   INT & (1335.2~\AA) & (3.2~\AA) & $  11$ &     0.03
 \\
\ion{C{\em l}}{1} & 1351.656 &  GAU1 & $+0.7{\pm}  0.2$ & $   26{\pm}   1$ & 
$    0.23{\pm}  0.00$ &     0.04 \\
\ion{Fe}{21} &  1354.08 &  GAU1 & $+5.6{\pm}  0.7$ & $  120{\pm}   2$ & 
$    0.52{\pm}  0.01$ &     0.04 \\
\ion{O}{1} & 1355.598 &  GAU1 & $ -2.1{\pm}  0.7$ & $   38{\pm}   1$ & 
$    0.15{\pm}  0.00$ &     0.04 \\
\ion{C}{1} & 1355.844 &  GAU1 & $+2.1{\pm}  0.7$ & $   38{\pm}   1$ & 
$    0.11{\pm}  0.00$ &     0.00 \\
\ion{O}{5} & 1371.296 &  GAU1 & $+6.0{\pm}  0.9$ & $   53{\pm}   2$ & 
$    0.12{\pm}  0.00$ &     0.05 \\
\ion{Si}{4} & 1393.760 & GAU2D & $+4.7{\pm}  0.2$ & $   47{\pm}   0$ & 
$     2.1{\pm}   0.0$ &     0.06 \\
\ion{Si}{4} & 1393.760 & GAU2D & $+14{\pm}   0$ & $  133{\pm}   1$ & 
$     2.9{\pm}   0.0$ & [$  0.5248,     -1.1$] \\
\ion{O}{4} & 1401.157 &  GAU7 & $+9{\pm}   1$ & $   91{\pm}   2$ & 
$    0.20{\pm}  0.00$ &     0.07 \\
\ion{O}{4} & 1397.202  & GAU7 & \nodata & \nodata &$   0.008{\pm} 0.003$
 &  \nodata   \\
\ion{O}{4} & 1399.769  & GAU7 & \nodata & \nodata &$   0.065{\pm} 0.004$
 &  \nodata   \\
\ion{O}{4} & 1404.780  & GAU7 & \nodata & \nodata &$   0.052{\pm} 0.004$
 &  \nodata   \\
\ion{S}{4} & 1404.808  & GAU7 & \nodata & \nodata &$   0.014{\pm} 0.001$
 &  \nodata   \\
\ion{S}{4} & 1406.051  & GAU7 & \nodata & \nodata &$   0.046{\pm} 0.004$
 &  \nodata   \\
\ion{O}{4} & 1407.379  & GAU7 & \nodata & \nodata &$   0.064{\pm} 0.004$
 &  \nodata   \\
\cutinhead{G160M}
\ion{Si}{4} & 1393.760 & GAU2D & $ -1.7{\pm}  0.2$ & $   60{\pm}   1$ & 
$     2.5{\pm}   0.0$ &     0.06 \\
\ion{Si}{4} & 1393.760 & GAU2D & $+4.9{\pm}  0.4$ & $  167{\pm}   2$ & 
$     2.6{\pm}   0.0$ & [$  0.552,      +7.1$] \\
\ion{O}{4} & 1401.157 &  GAU7 & $+10{\pm}   1$ & $   92{\pm}   3$ & 
$    0.21{\pm}  0.01$ &     0.07 \\
\ion{O}{4} & 1397.202  & GAU7 & \nodata & \nodata &$   0.008{\pm} 0.003$
 &  \nodata   \\
\ion{O}{4} & 1399.769  & GAU7 & \nodata & \nodata &$   0.055{\pm} 0.004$
 &  \nodata   \\
\ion{O}{4} & 1404.780  & GAU7 & \nodata & \nodata &$   0.038{\pm} 0.004$
 &  \nodata   \\
\ion{S}{4} & 1404.808  & GAU7 & \nodata & \nodata &$   0.014{\pm} 0.001$
 &  \nodata   \\
\ion{S}{4} & 1406.051  & GAU7 & \nodata & \nodata &$   0.046{\pm} 0.004$
 &  \nodata   \\
\ion{O}{4} & 1407.379  & GAU7 & \nodata & \nodata &$   0.063{\pm} 0.005$
 &  \nodata   \\
\ion{Si}{2} & 1526.707 &  GAU1 & $+0.2{\pm}  0.4$ & $   50{\pm}   1$ & 
$    0.60{\pm}  0.01$ &     0.10 \\
\ion{Si}{2} & 1533.431 &  GAU1 & $+3.2{\pm}  0.4$ & $   51{\pm}   1$ & 
$    0.67{\pm}  0.01$ &     0.10 \\
\ion{C}{4} & 1548.204 & GAU2D & $+2.6{\pm}  0.2$ & $   48{\pm}   0$ & 
$     4.2{\pm}   0.1$ &     0.14 \\
\ion{C}{4} & 1548.204 & GAU2D & $+18{\pm}   0$ & $  151{\pm}   1$ & 
$     6.8{\pm}   0.1$ & [$  0.564,      +3.3$] \\
\ion{He}{2} & 1640.404 &  GAU2 & $+4.1{\pm}  0.2$ & $   50{\pm}   1$ & 
$     5.9{\pm}   0.1$ &     0.15 \\
\ion{He}{2} & 1640.404 &  GAU2 & $+20{\pm}   1$ & $  128{\pm}   2$ & 
$     4.5{\pm}   0.1$ &  \nodata  \\
\ion{C}{1} & 1657{\em m} &   INT & (1657.1~\AA) & (3.0~\AA) & $     5.8$ & 
    0.21 \\
\ion{A{\em l}}{2} & 1670.787 &  GAU2 & $+1.9{\pm}  0.7$ & $   33{\pm}   3$ & 
$    0.52{\pm}  0.07$ &     0.18 \\
\ion{A{\em l}}{2} & 1670.787 &  GAU2 & $ -1{\pm}   1$ & $  107{\pm}   7$ & 
$     1.0{\pm}   0.1$ &  \nodata  
\enddata
\tablecomments{See comments table 3.}
\end{deluxetable}

\clearpage
\begin{deluxetable}{llccccc}
\tabletypesize{\footnotesize}
\tablenum{5} 
\tablecaption{{\em HST}\/ Measurements: COS Visit 2 (Flare)} 
\tablecolumns{7}
\tablewidth{0pt} 
\tablehead{
\colhead{Species} & \colhead{$\lambda_{\rm lab}$} & \colhead{Flag} & \colhead{$\upsilon_{\rm L}$} & \colhead{FWHM$_{\rm L}$} & 
\colhead{$f_{\rm L}$} & \colhead{$f_{\rm C}$}\\
\colhead{} & \colhead{(\AA)} & \colhead{} & \multicolumn{2}{c}{(km s$^{-1}$)} &  
\colhead{(10$^{-14}$ cgs)} &
\colhead{(10$^{-14}$ cgs \AA$^{-1}$)}\\
\colhead{(1)} & \colhead{(2)} & \colhead{(3)} & \colhead{(4)} & \colhead{(5)} & \colhead{(6)} & 
\colhead{(7)} 
}
\startdata
\ion{Si}{4} & 1393.760 & GAU2D & $+0.8{\pm}  0.6$ & $   61{\pm}   2$ & 
$     9.1{\pm}   0.3$ &     0.39 \\
\ion{Si}{4} & 1393.760 & GAU2D & $+31{\pm}   2$ & $  207{\pm}   4$ & 
$     9.6{\pm}   0.3$ & [$  0.562,      +6.2$] \\
\ion{O}{4} & 1401.157 &  GAU7 & $+19{\pm}   3$ & $  124{\pm}   8$ & 
$     1.1{\pm}   0.1$ &     0.39 \\
\ion{O}{4} & 1397.202  & GAU7 & \nodata & \nodata &$   0.07{\pm} 0.03$
 &  \nodata   \\
\ion{O}{4} & 1399.769  & GAU7 & \nodata & \nodata &$    0.45{\pm}  0.04$
 &  \nodata   \\
\ion{O}{4} & 1404.780  & GAU7 & \nodata & \nodata &$    0.25{\pm}  0.04$
 &  \nodata   \\
\ion{S}{4} & 1404.808  & GAU7 & \nodata & \nodata &$   0.06{\pm} 0.01$
 &  \nodata   \\
\ion{S}{4} & 1406.051  & GAU7 & \nodata & \nodata &$    0.20{\pm}  0.04$
 &  \nodata   \\
\ion{O}{4} & 1407.379  & GAU7 & \nodata & \nodata &$    0.42{\pm}  0.04$
 &  \nodata   \\
\ion{Si}{2} & 1526.707 &  GAU1 & $ 0{\pm}   1$ & $   53{\pm}   3$ & 
$     1.5{\pm}   0.1$ &     0.53 \\
\ion{Si}{2} & 1533.431 &  GAU1 & $+1{\pm}   2$ & $   58{\pm}   4$ & 
$     1.6{\pm}   0.1$ &     0.53 \\
\ion{C}{4} & 1548.204 & GAU2D & $+5.7{\pm}  0.4$ & $   45{\pm}   1$ & 
$  26{\pm}   0$ &     0.58 \\
\ion{C}{4} & 1548.204 & GAU2D & $+41{\pm}   1$ & $  197{\pm}   2$ & 
$  43{\pm}   0$ & [$  0.583,     -0.2$] \\
\ion{He}{2} & 1640.404 &  GAU2 & $+6.8{\pm}  0.4$ & $   50{\pm}   1$ & 
$  26{\pm}   1$ &     0.69 \\
\ion{He}{2} & 1640.404 &  GAU2 & $+36{\pm}   2$ & $  164{\pm}   5$ & 
$  18{\pm}   1$ &  \nodata  \\
\ion{C}{1} & 1657{\em m} &   INT & (1657.1~\AA) & (3.0~\AA) & $  13$ &     0.74
 \\
\ion{A{\em l}}{2} & 1670.787 &  GAU2 & $ -2{\pm}   1$ & $   27{\pm}   4$ & 
$     1.6{\pm}   0.2$ &     0.77 \\
\ion{A{\em l}}{2} & 1670.787 &  GAU2 & $+19{\pm}   5$ & $  169{\pm}  14$ & 
$     4.4{\pm}   0.3$ &  \nodata 
\enddata
\tablecomments{See comments Table~3.}
\end{deluxetable}

\clearpage
Reference (``laboratory'') wavelengths (in vacuum) were taken from Ayres (2015), except for the solar flare line [\ion{Fe}{21}], which was not included in the $\alpha$~Cen study.  The NIST Atomic Spectra Database (ASD: Kramida et al.\ 2013)\footnote{see: http://www.physics.nist.gov/PhysRefData/ASD/lines\_form.html} wavelength (1354.08~\AA) was adopted for this key coronal forbidden transition, but the quality of the measurement is rated poor.  Recently, Young et al.\ (2015) proposed a revised wavelength of 1354.106${\pm}$0.023~\AA, based on measurements of [\ion{Fe}{21}] in solar flares.  If correct, it would shift the [\ion{Fe}{21}] velocities in Tables~3-5 by $6{\pm}5$~km s$^{-1}$ to the blue.

A variety of fitting strategies were employed, following the lead of the Ayres (2015) study of STIS FUV echelle spectra of $\alpha$~Cen A and B.  The particular measurement scenario is flagged explicitly in col.~3 of the Tables, as follows.  ``INT'' refers to a numerical flux integration over the velocity bandpass listed in col.~5.  The value in col.~7 is a derived continuum flux density, assumed constant over the spectral feature or multiplet, based on filtering the surrounding intensities to suppress emission lines.  ``GAU1'' refers to a single Gaussian fit.  ``GAU1S'' is a special case in which a sloping continuum was allowed: col.~7 now contains the zero point and gradient of a linear continuum fit: $(f_{\rm C})_{\lambda}= C_{0} + C_{1}\,(\lambda - \lambda_{\rm L})$, where $f_{\rm C}$ is the monochromatic continuum flux density and $\lambda_{\rm L}$ is the transition wavelength.  ``GAU1X'' is another specialized fit, applied to the ISM absorption core of \ion{H}{1} Ly$\alpha$.  The fitting function is of the form $e^{-x^4}$, where $x\equiv \Delta\lambda/\Delta\lambda_{D}$ and $\Delta\lambda_{D}$ is the ``Doppler width.''  Compared with the normal Gaussian ($e^{-x^2}$), the steep-sided flat-bottomed GAU1X profile is more appropriate for the saturated Ly$\alpha$ absorption core.  However, the FWHM and total flux have little meaning, especially because the imposed continuum level (at the apparent peak of Ly$\alpha$) is rather arbitrary, so the derived values are not reported in the table; only the velocity shift is significant.

``GAU2'' is a bimodal Gaussian fit of the narrow/broad variety, used specifically for the \ion{N}{5}\,+\,\ion{Si}{4}\,+\,\ion{C}{4} collective filtered profiles, but also for other isolated features such as \ion{Si}{3} $\lambda$1206.  For the filtered hot-line average, the component fluxes are narrow/total and broad/total.  ``GAU2R'' refers to a double-Gaussian fit applied to ostensibly unrelated features (say, stellar emission and superimposed interstellar absorption), for which the second component is negative in flux to mimic a central reversal or ISM absorption.  In these cases, the width of the principal component should not be taken too literally, although the velocity shift should be reliable.

``GAU7'' is a 7-component Gaussian model covering combined multiplets of similar transitions, namely \ion{O}{4}]\,+\,\ion{S}{4}], for which the widths of the features were constrained to be the same, and the Doppler shifts as well (i.e., the wavelength differences between the components were fixed at the reference values, but a shift of the whole transition array was allowed).  The derived collective velocity shift and average profile width, together with the associated uncertainties (and the continuum flux) are listed for the principal transition.  For the remaining multiplet members, only the unique flux values are provided.

Finally, ``GAU2D'' refers to a specialized constrained bimodal Gaussian fit applied to the hot-line doublets, such as \ion{Si}{4} $\lambda$1393\,+\,$\lambda$1402.  Figure~10 illustrates such fits to the (un-smoothed) EK~Dra hot lines from the co-added STIS spectrum.  This bimodal strategy for decomposing the complex hot-line profile shapes of cool stars has a long history, as discussed recently by Ayres (2015) in the context of low-activity $\alpha$~Cen A (and companion B [K1~V]), and for the earlier COS pointing on EK~Dra itself by Ayres \& France (2010).  The upper paired panels of Fig.~10 illustrate the application to the \ion{N}{5}, \ion{Si}{4}, and \ion{C}{4} resonance doublets: principal components to the left, secondary to the right (but with one-half the $y$-axis scale).  Under optically-thin conditions, the weaker components would be half the flux densities of the stronger ones, and thus should appear similar in these scaled diagrams.  That expectation is met for \ion{N}{5}, but becomes progressively out of sync for \ion{Si}{4} and \ion{C}{4}.  The fitting procedure assumed that the doublet members were self-similar: NC of the same width, BC of the same width;  Doppler shifts of the NC the same, and similarly for the BC (although possibly different between narrow and broad); and the same flux ratios, narrow/broad.  Only the relative total fluxes between the doublet members was allowed to float, as well as the wavelength difference between the two transitions, to allow for small errors in the laboratory values, or subtle residual distortions in the STIS wavelength scales.

\begin{figure}
\figurenum{10}
\vskip  0mm
\hskip  14mm
\includegraphics[scale=0.75]{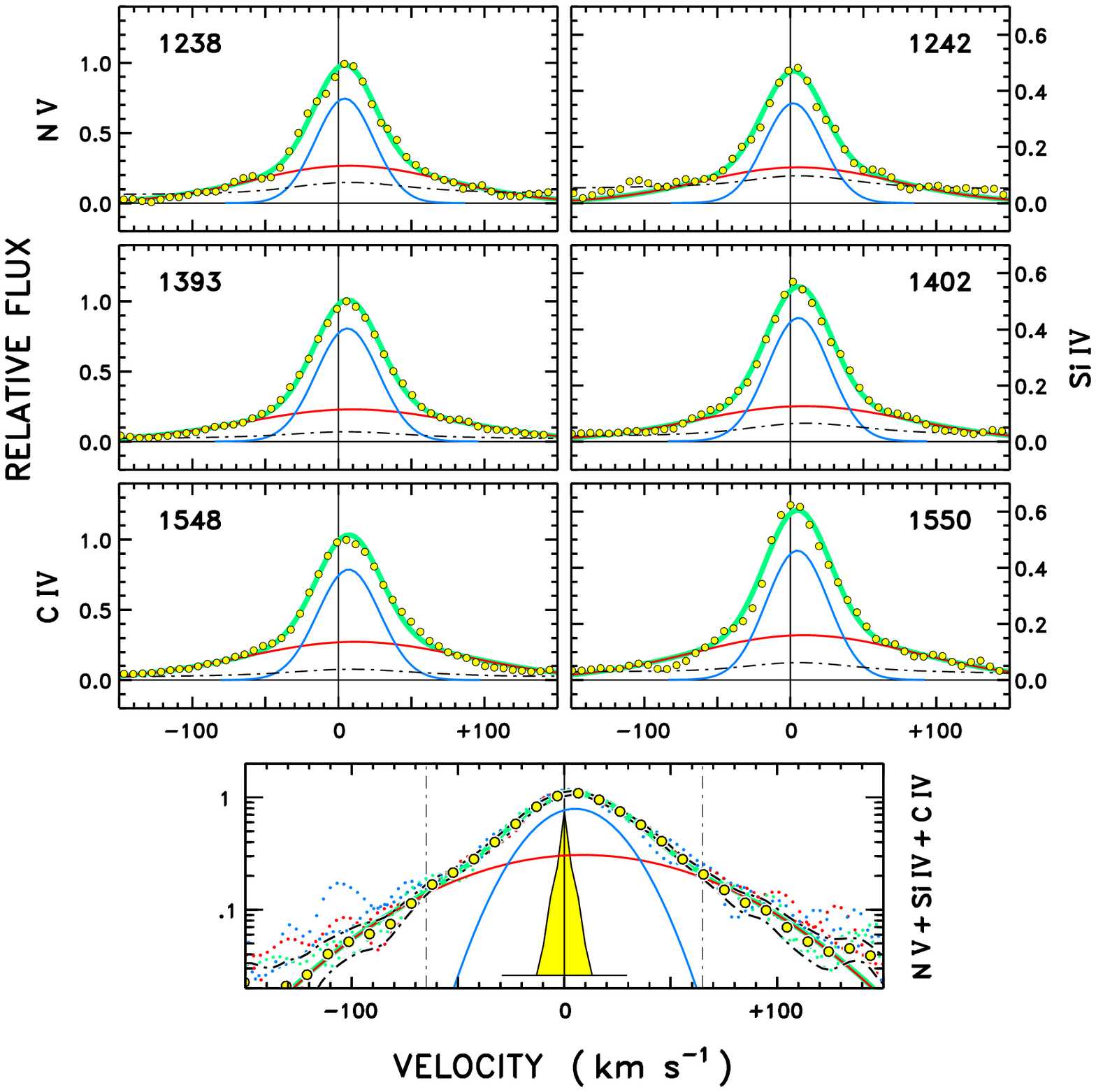} 
\vskip -3mm
\figcaption[]{\small Bimodal Gaussian fits to the EK~Dra hot-line doublets, and a filtered average of the three, from the co-added STIS spectrum.  In the upper paired panels, black dotted-dashed curves are smoothed 1\,$\sigma$ photometric errors (per resel).  In the bottom panel, the black dotted dashed curves indicate ${\pm}3$~standard errors of the mean (s.e.) of the filtered average (per bin).  The filtering was designed to suppress accidental emission blends from unrelated species.  The small s.e.'s demonstrate that the profiles are close to self-similar, aside from the occasional blend.  Open circles represent the STIS spectra: only every other point is plotted. In the upper panels, the two members of each doublet were fitted simultaneously, imposing various constraints on the narrow and broad sub-component widths, shifts, and strengths (see text).  The NC are blue, the BC are red, and the sum is green.  The STIS E140M line spread function for the photometric aperture (FWHM$\sim$~8 km s$^{-1}$: yellow shaded profile in the bottom panel) was accounted in the modeling procedure.  Note the factor of 2 $y$-axis scale change between the upper left panels, for the principal doublet members, and the right panels, for the secondary members.  With that scaling, the two features should have similar appearance in the optically thin limit.
}
\end{figure}

For this GAU2D case, the first set of parameters listed (in Tables 3-5) is for the NC of the constrained doublet fit, and col.~7 holds the continuum flux as usual.  The col.~2 wavelength is for the principal member of the doublet. The second set of parameters listed is for the BC.  However, highlighted by square brackets, col.~7 now has the doublet flux ratio, weaker/stronger, which should be close to 0.5 in the optically thin limit (for the specific hot-line doublets here); as well as the deviation of the fitted wavelength difference from the reference value, expressed in km s$^{-1}$.  These latter shifts are small for the STIS spectrum, but can be substantial in several of the COS cases, mainly reflecting residual errors in the simple correction of the COS wavelengths.

The bottom panel of Fig.~10 illustrates a GAU2 bimodal fit to a filtered sum of the three hot-line doublets, each member (six altogether) scaled to the same integrated flux over the interval $|\upsilon| \le 65$~km s$^{-1}$.  The scaled fluxes were combined on a common velocity grid, applying an Olympic filter (throwing out the highest and lowest of the six values) for the central ${\pm}65$~km s$^{-1}$ zone, and a more aggressive filtering (eliminating the three highest values) for the outer intervals where the influence of weak emission blends on the broad component is strongest.  As with the individual bimodal fits, the hybrid profile has a relatively narrow NC, about 50~km s$^{-1}$ FWHM, and is redshifted by 5~km s$^{-1}$, similar to those of low-activity G dwarfs like $\alpha$~Cen A; but a very broad BC, about three times wider than the NC, and more redshifted (9~km s$^{-1}$).  The enhanced broadening and redshift of EK~Dra's BC are the main distinguishing characteristics compared with low-activity counterparts like $\alpha$~Cen A (e.g., Fig.~2).

Uncertainties on the fitted parameters were estimated by a Monte Carlo approach: a reference profile (the derived fit itself) was perturbed by random realizations of the smoothed photometric error, and then refitted.  The standard deviations of the derived parameters over many such trials were taken as a fair gauge of the measurement uncertainties devolving from the photometric noise component.

\subsection{The Isolated FUV Continuum Bursts}

Recall the two bursts (at $t= $3.6~hr and 3.7~hr) in the G160M blue and red continuum bands, which remarkably were not mirrored in any of the other activity tracers.  These isolated bursts are distinctly bluer (indicating a hotter continuum) than the more equal count rates of the two separated continuum bands later on, in quiet visits 3-5.  This continuum-only behavior is challenging to understand, and seems to have few, if any, precedents among existing records of stellar flare events.  This likely is because there are only a handful of stellar flare campaigns simultaneously at UV and soft X-ray energies.  One example is Mitra-Kraev et al.\ (2005), who were able to use the Optical Monitor on {\em XMM-Newton,}\/ with broad-band FUV and NUV filters, to jointly record a number of UV/X-ray events on several dMe stars.  The authors found that the UV rise typically preceded the soft X-rays by several minutes, which usually is the case for white-light flares (WLF) on the Sun.  There is perhaps only a single example in their several time series where a UV burst was seen during a flare decay without an X-ray counterpart (their Fig.~1).  On the Sun,  WLFs are difficult to separate from the glare of the photosphere, but if one is seen, it usually will have an extensive supporting context, especially at high energies.  Fang \& Ding (1995) divided solar WLFs into two classes: Type~I, or ``prompt,'' events which are coincident with the hard X-ray (HXR) burst, and apparently result from direct excitation of the chromospheric gas by the flare's energetic particle beams; and the less frequent Type~II, or ``delayed,'' bursts, which occur several minutes after the HXR impulsive phase, and appear to be deeper seated, in the photosphere itself.  Fang \& Ding did not speculate on a source for the Type~IIs.  Perhaps the delayed FUV continuum brightenings on EK~Dra are related to the solar WLF Type~IIs.  One idea, motivated by the hot-line redshifts during the flare decay, is that the bursts are a response to the ballistic splash-down of the cooling gas onto the photospheric layers; something analogous to accretion streams in T-Tauri stars.

\subsection{The [\ion{Fe}{21}] Coronal Forbidden Line}

Ayres et al.\ (2003) carried out a search for coronal forbidden lines in STIS FUV spectra of the late-type objects with sufficiently deep exposures available at the time.  In the hyper-active members of the sample, like RS~Canum Venaticorum binaries and dMe flare stars, the intrinsically faint [\ion{Fe}{21}] $\lambda$1354 feature often was visible.  Furthermore, the [\ion{Fe}{21}] line strength was found to be linearly correlated with the {\em ROSAT}\/ soft X-ray flux.  In the moderate-to-slow rotators (in terms of $v\sin{i}$), the average FWHM of the coronal iron line was about 110~km s$^{-1}$, somewhat larger than the expected thermal width of $\sim 90$~km s$^{-1}$ at the formation temperature of 10~MK.  In several of the fast rotators, there was a suggestion of $\sim2{\times}$ enhanced broadening beyond that expected from $v\sin{i}$ alone.  The ``super-rotational'' broadening was attributed to highly extended hot coronal structures.

In EK~Dra, the COS visits 3-5 average profile of [\ion{Fe}{21}] does not display obviously extended red wings, like the lower temperature subcoronal features, although certainly the coronal forbidden line is very broad, approximately mid-way in width between the hot-line NC and BC.  Nevertheless, the FWHM$\sim 120$~km s$^{-1}$ is close enough to the low-$v\sin{i}$ average noted above that it would be difficult to make a case for super-rotational broadening.

The COS velocity of [\ion{Fe}{21}] in the visits 3-5 average spectrum is $+6$~km s$^{-1}$ for the NIST reference wavelength, although with the Young et al.\ (2015) revised value, the shift would decrease to about 0 ($\pm 5$~km s$^{-1}$ from the $\sigma_{\lambda}$ of the reference wavelength).  At the same time, the STIS velocity of [\ion{Fe}{21}] would shift to $-10$~km s$^{-1}$ ($\pm 6$~km s$^{-1}$, where the larger $\sigma$ partly is due to the reference wavelength, and partly to the low S/N of the measurement).

Because [\ion{Fe}{21}] correlates well with the {\em ROSAT}\/ flux for hyper-active coronal sources, the measured intensity during the pre- and post-flare intervals can be exploited to estimate the quiescent X-ray flux of EK~Dra during the 2012 campaign, noting that the existing {\em ROSAT}\/ survey $L_{\rm X}= 10^{29.9}$ erg s$^{-1}$ (0.1-2.4~keV) reported by G\"udel et al.\ (1995) is not exactly simultaneous.  The scaling relation of Ayres et al.\ (2003) can be expressed as:
\begin{equation}
L_{\rm X}\sim 2{\times}10^{3}\,(\frac{f_{\rm Fe XXI}}{f_{\rm bol}})\,L_{\rm bol}
\end{equation}
The stellar bolometric flux, $f_{\rm bol}\sim 2.45{\times}10^{-8}$ erg cm$^{-2}$ s$^{-1}$, can be estimated from $V$ and a bolometric correction derived from $B-V$ (see Ayres et al.\ 2005).  The bolometric luminosity, $L_{\rm bol}\sim 3.4{\times}10^{33}$ erg s$^{-1}$, follows from the distance (34~pc), and is similar to $L_{\odot}$.  The [\ion{Fe}{21}] flux ratio, based on the STIS/COS average $f_{\rm L}\sim 6{\times}10^{-15}$ erg cm$^{-2}$ s$^{-1}$, is $2.5{\times}10^{-7}$: in the upper tier among the coronal objects considered by Ayres et al.\ (2003).  Finally, the inferred X-ray luminosity is $L_{\rm X}\sim 1.7{\times}10^{30}$ erg s$^{-1}$, about twice the {\em ROSAT}\/ survey value, but certainly consistent within the uncertainties of the [\ion{Fe}{21}] scaling law, and the temporal variability of $L_{\rm X}$ noted by G{\"u}del et al.\ (1995).

\begin{figure}
\figurenum{11}
\vskip  0mm
\hskip  10mm
\includegraphics[scale=0.75]{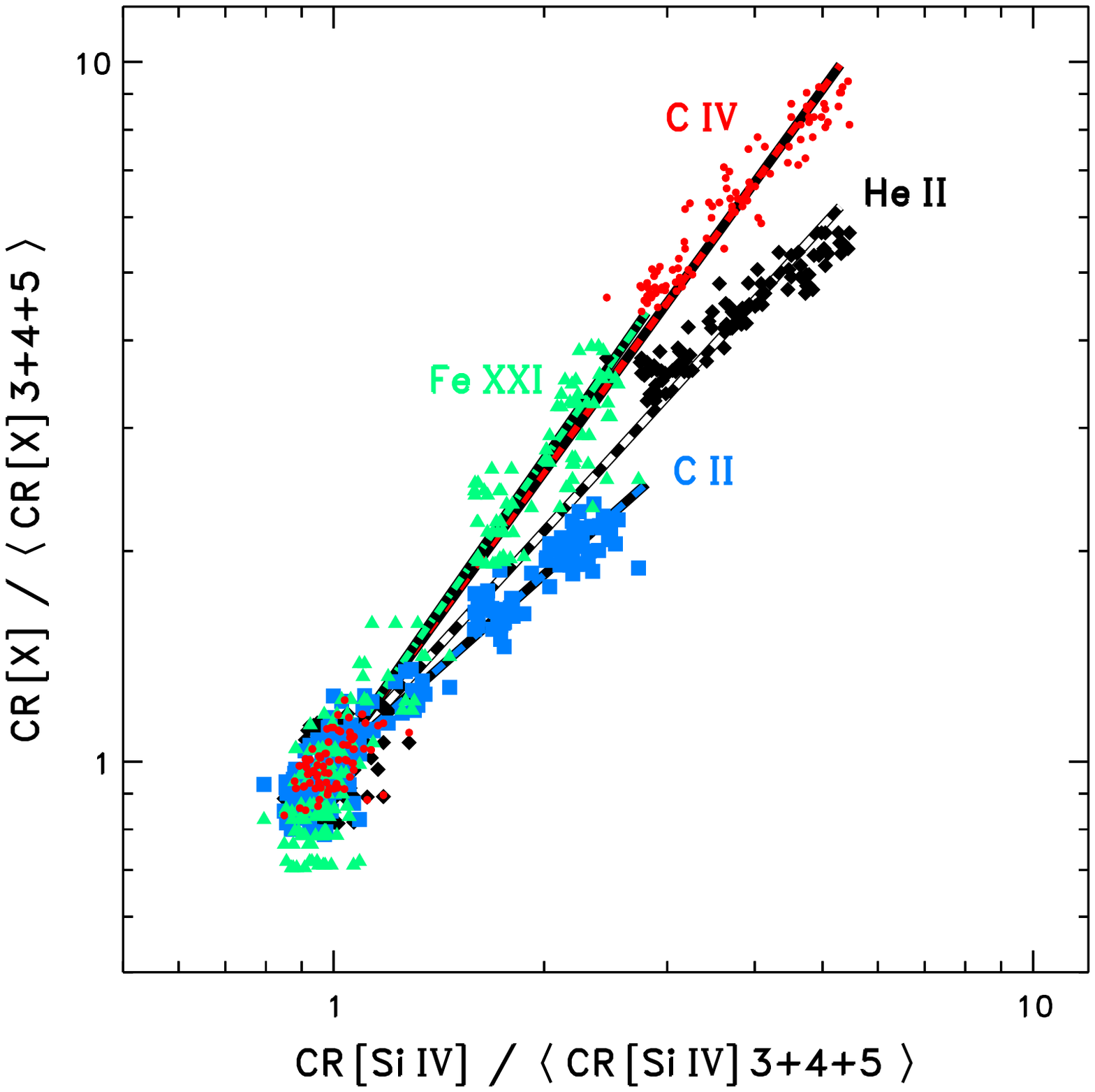} 
\vskip 0mm
\figcaption[]{\small Correlation diagram pitting various diagnostics from all the visits, normalized to the visits 3-5 average, against the similarly normalized \ion{Si}{4} doublet count rates.  The systematic trends toward the upper right represent the flare decay in visit 2, while the dispersions of the clusters of points near the origin (1,1) indicate the typical variability during quieter times.
}
\end{figure}

\subsection{Correlations among Bright Lines in the Flare and Quiescent Periods}

Figure~11 is a correlation diagram pitting various diagnostics against the \ion{Si}{4} doublet, which was recorded in every COS observation.  As with the previous diagram(s), the individual flux values from all the visits were normalized to the corresponding averages over visits 3-5.  The upper distributions of symbols are contributed by the large flare, while the clump at (1,1) is dominated by the quiescent intervals, which nevertheless display low-level variability including a small flare decay.  The straight lines are power-law fits to the individual distributions.  The derived slopes range from 1.44 and 1.38 for [\ion{Fe}{21}] and \ion{C}{4}, respectively, to slightly above unity for \ion{He}{2} (1.10), and slightly below for chromospheric \ion{C}{2} (0.89).  These slopes mirror similar power laws seen between, say, X-rays and \ion{C}{4} in diverse samples of solar-type dwarfs (e.g., Ayres et al.\ 1995), and in this instance must reflect the time-evolution of the post-flare cooling process.

\subsection{Quiescent and Flare Profiles of \ion{C}{4}}

Figure~12 compares \ion{C}{4} doublet profiles from the COS G160M exposures of the visit~2 flare decay, the visits 3-5 average COS G160M spectrum, and a multi-epoch STIS average for the quiet solar twin $\alpha$~Cen A (Ayres 2015); all including bimodal constrained Gaussian fits.  The quiescent profiles of EK~Dra show redshifts of both components, less for the NC ($\sim +6$~km s$^{-1}$, for the average of the STIS \ion{C}{4} doublet members), more for the BC ($\sim +10$~km s$^{-1}$, for the STIS doublet average).  This compares with a somewhat smaller redshift of both $\alpha$~Cen A components ($\sim +5$ km s$^{-1}$ for each).  The width of the EK~Dra \ion{C}{4} NC in the quiet spectrum (50~km s$^{-1}$) is somewhat larger than in $\alpha$~Cen A  (36~km s$^{-1}$), but the BC is much broader (180~km s$^{-1}$ vs.\ 66~km s$^{-1}$).  In the flare, remarkably, the \ion{C}{4} NC has a similar redshift and FWHM ($+6$~km s$^{-1}$ and 45~km s$^{-1}$, respectively) to the ``quiet'' average, and indeed to the low-activity solar twin $\alpha$~Cen A.  At the same time, the flare BC not only is broader than in the quiet cases ($\sim 200$~km s$^{-1}$), but is substantially more redshifted (at about +40~km s$^{-1}$).  The larger BC redshifts during the flare are notable but not unusual: similar behavior, with similar velocity shifts, has been seen previously in STIS spectra of flares on the dMe stars AU~Mic (Robinson et al.\ 2001) and AD~Leo (Hawley et al.\ 2003), and appears to be a signature of the post-flare cooling process.   What perhaps is more remarkable is that despite the very large difference in emission levels of the two solar analogs (expressed, say, in terms of surface fluxes), the ``quiet'' \ion{C}{4} profiles are as similar as they are, not only in the bimodal nature of the line shape asymmetries, but also in the magnitudes of the velocity shifts and line widths.

\begin{figure}
\figurenum{12}
\vskip  0mm
\hskip  -5mm
\includegraphics[scale=0.775,angle=90]{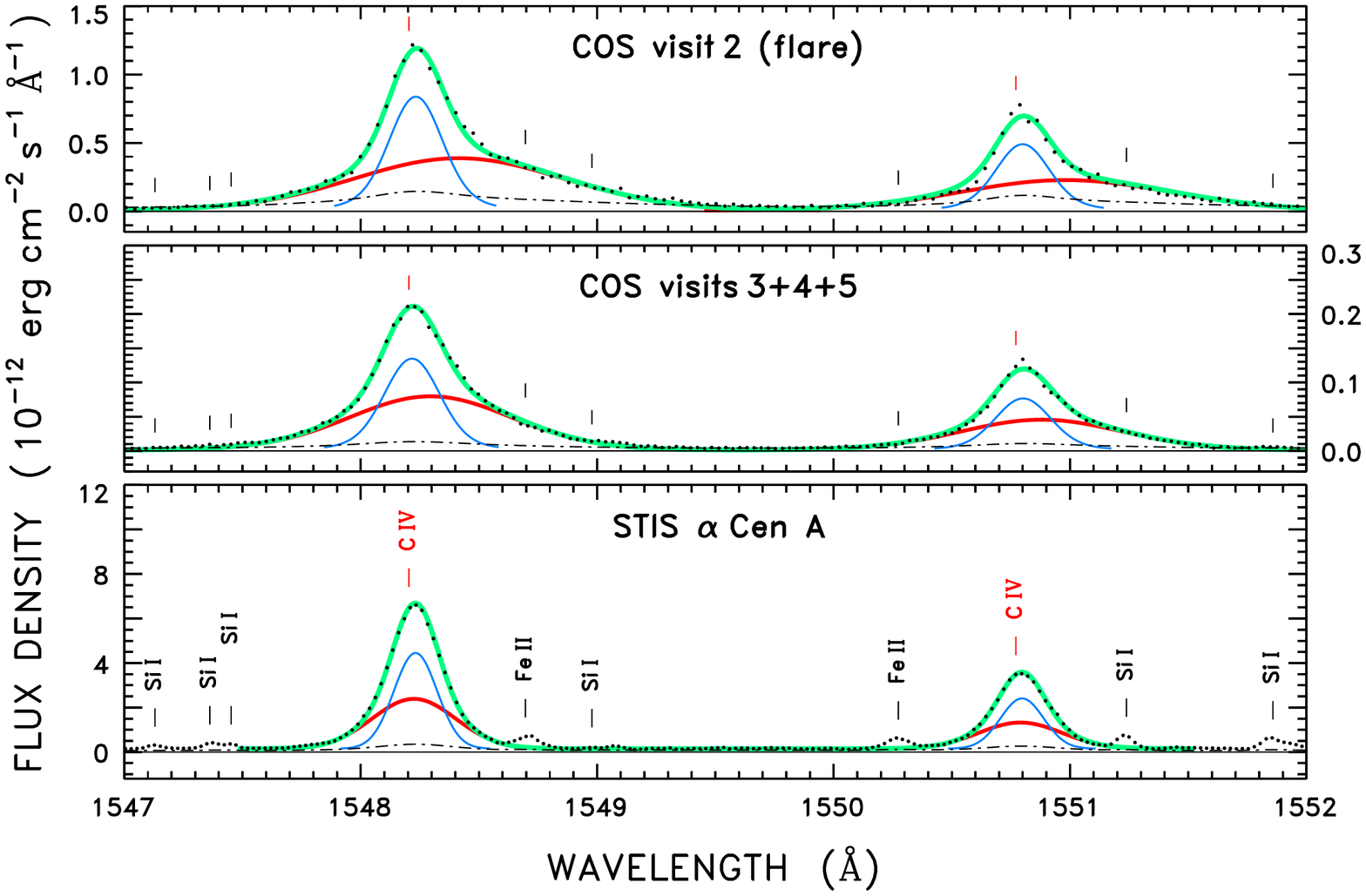} 
\vskip 0mm
\figcaption[]{\small \ion{C}{4} doublet profiles, and bimodal fits, from the EK~Dra flare (COS visit~2: upper panel) and the quiet average (COS visits 3-5: middle panel), compared to the low-activity sunlike star $\alpha$~Cen A (from STIS: lower panel).  Only every third point (black dots) is plotted for these oversampled broad profiles.  Note the large scale change between the upper (flare) and middle (quiescent) panels for EK~Dra.  Black dotted-dashed curves are smoothed 10\,$\sigma$ photometric errors (per resel).  The EK~Dra flare profiles are remarkably similar to the quiet average, despite the large flux increase, except for the more redshifted broad components.  The \ion{C}{4} NC of EK~Dra are somewhat similar to those of $\alpha$~Cen A, but the BC are much broader and occupy a larger fraction of the total doublet flux.
}
\end{figure}

\begin{figure}
\figurenum{13}
\vskip  0mm
\hskip  -5mm
\includegraphics[scale=0.75,angle=90]{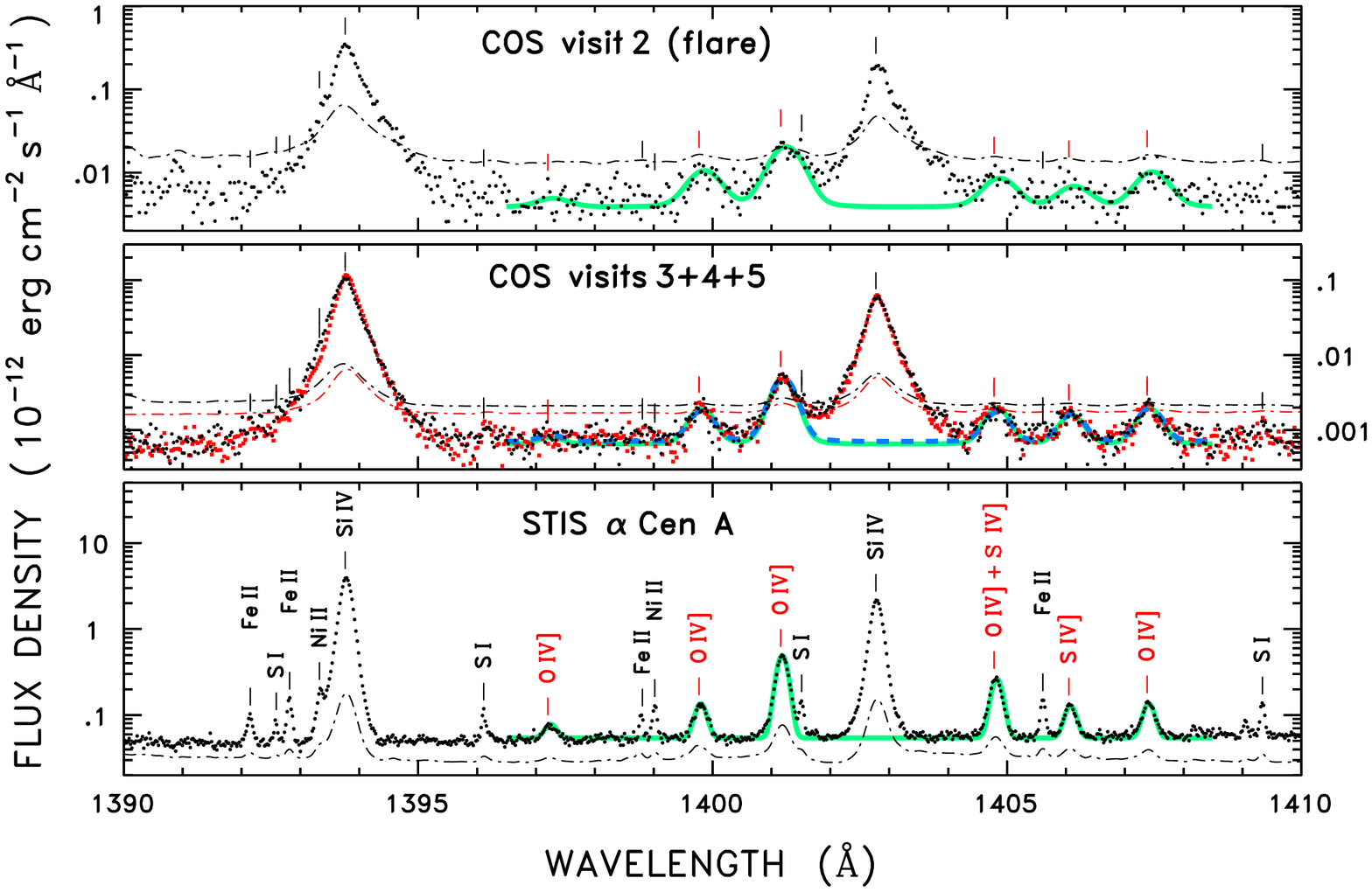} 
\vskip 0mm
\figcaption[]{\small Density diagnostic \ion{O}{4}] (plus \ion{S}{4}]) 1400~\AA\ multiplet, together with multi-component Gaussian model (thick curves).  Same arrangement of spectra as in Fig.~12.  Note the logarithmic $y$-axes, and large scale change between the two EK~Dra panels.  As with the previous figure, only every third point (dots) is plotted.  Dotted-dashed curves again are smoothed 10\,$\sigma$ photometric errors.  Note also that the middle panel contains both G130M (red squares for data; green curve for fit; red dotted-dashed curve for 10\,$\sigma$ level) and G160M (black dots; blue dashed; black dotted-dashed) for the visits 3-5 average spectra, while the upper panel has only G160M (from the initial flare interval in visit 2).  In the G130M/G160M comparison, the agreement between the wavelength-corrected profiles of \ion{Si}{4} $\lambda$1402 is excellent, while the shorter wavelength component, $\lambda$1393, displays a noticeable blue deviation in G160M, as noted earlier in Fig.~8.  The numerous narrow lines of chromospheric species (e.g., \ion{S}{1}, \ion{Fe}{2}, and \ion{Ni}{2}) seen in low-activity $\alpha$~Cen A are expected to be much less prominent in hyper-active EK~Dra, owing to the general decline in importance of the lower temperature layers, relative to the hotter subcoronal plasma, with increasing activity.
}
\end{figure}

\begin{deluxetable}{lcccc} 
\tabletypesize{\footnotesize}
\tablenum{6} 
\tablecaption{\ion{O}{4}] Line Ratios} 
\tablecolumns{5}
\tablewidth{0pt} 
\tablehead{
\colhead{Dataset} & \colhead{$R_{\rm 1}\ast$} & \colhead{$R_{\rm 2}\ast$} & \colhead{}  &  \colhead{} \\[5pt]
\colhead{} & \colhead{$\frac{{\scriptsize \onehalf}({\lambda}1399\,+\,{\lambda}1407)}{{\lambda}1401}$} & 
\colhead{$\frac{{\scriptsize \onehalf}({\lambda}1399\,+\,{\lambda}1407)}{{\lambda}1404}$} & 
\colhead{$\frac{{\lambda}1399}{{\lambda}1407}$} & 
\colhead{$\frac{{\lambda}1397}{{\lambda}1404}$}\\[5pt]
\colhead{(1)} & \colhead{(2)} & \colhead{(3)} & \colhead{(4)} & \colhead{(5)}
}
\startdata
{Theoretical Limits}        & 0.16--0.42  & 0.3--2.5  & 1.02  & 0.13  \\[5pt]
{Electron Density Range (cm$^{-3}$)}        & $10^{10}-10^{12}$ &
                              $10^{10}-10^{12}$ &
                              \multicolumn{2}{c}{$\ne n_{\rm e}$} \\[5pt]
\cutinhead{EK~Draconis}
{Quiet Average: COS G130M}  & $0.32{\pm}0.02$ & $1.2{\pm}0.1$ & $1.0{\pm}0.1$ & $0.15{\pm}0.05$ \\
{Quiet Average: COS G160M}  & $0.28{\pm}0.02$ & $1.6{\pm}0.2$ & $0.9{\pm}0.1$ & $0.21{\pm}0.08$ \\
{Flare: COS G160M}          & $0.40{\pm}0.04$ & $1.7{\pm}0.3$ & $1.1{\pm}0.1$ & $0.27{\pm}0.11$ \\
\cutinhead{$\alpha$~Centauri A} 
{Quiet Average: STIS E140M} & $0.185{\pm}0.002$ & $0.429{\pm}0.006$ & $0.99{\pm}0.03$ & $0.127{\pm}0.007$ \\
\enddata
\tablecomments{Theoretical values from Keenan et al.\ (2009).}
\end{deluxetable}

\subsection{Densities from the \ion{O}{4}] Semi-Forbidden Multiplet}

Flower \& Nussbaumer (1975) originally proposed that flux ratios within the \ion{O}{4}] 1400~\AA\ intercombination multiplet are sensitive to density over the range $n_{\rm e}\sim 10^{10}-10^{12}$ cm$^{-3}$ thought to be characteristic of solar conditions -- quiescent at the low end, flaring on the high side -- at the O$^{3+}$ ionization equilibrium peak temperature of $\sim 10^5$~K.  There is a long history of applications to solar and stellar plasmas (see, e.g., Cook et al.\ 1995, and references therein), continuing to the present (e.g., Keenan et al.\ 2009; Ayres 2015).  Figure~13 illustrates the \ion{O}{4}] region from the initial visit~2 flare interval (G160M, only); the quiet averages (visits 3-5) from both G160M and G130M; and the $\alpha$~Cen A STIS spectrum.  The logarithmic presentation emphasizes the wider, more redshifted broad components of the EK~Dra profiles compared with low-activity $\alpha$~Cen.  The 7-component Gaussian fits (five \ion{O}{4}], two \ion{S}{4}] transitions) take into account the small contamination of \ion{O}{4}] $\lambda$1404 by \ion{S}{4}] $\lambda$1404, scaled from \ion{S}{4}] $\lambda$1406 (one of the fitted features) according to plasma emission models (see Ayres 2015).

Table~6 summarizes \ion{O}{4}] diagnostic ratios constructed from the COS measurements in Tables~4 and 5, and including values from the multi-epoch average $\alpha$~Cen A STIS spectrum.  The nomenclature for the ratios is that of Ayres (2015): the numerators in the traditional density-sensitive $R_{1}\equiv {\lambda}1407 / {\lambda}1401$ and $R_{2}\equiv {\lambda}1407 / {\lambda}1404$ were replaced by $\onehalf({\lambda}1399 + {\lambda}1407)$ to improve S/N with these generally faint emissions: ${\lambda}1399$ and ${\lambda}1407$ share a common upper level and their ratio is very close to unity according to theory.  The hybrid numerator ratios are designated $R\ast$.  The ratio ${\lambda}1397 / {\lambda}1404$ also is composed of transitions that share a common upper level, with a theoretical value of 0.13 (Keenan et al.\ 2009).  However, in this case there is no advantage to form a hybrid flux between ${\lambda}1397$ and ${\lambda}1404$ because the former transition usually is too low in S/N, if detected at all, in normal astrophysical sources.

The $\alpha$~Cen ${\lambda}1399 / {\lambda}1407$ and ${\lambda}1397 / {\lambda}1404$ ratios are close to the theoretical expectations.  This also is the case for EK~Dra, although the uncertainties are larger, especially for marginally detected ${\lambda}1397$.  Note, however, that the diagnostic ratios $R_{1}\ast$ and $R_{2}\ast$ are significantly elevated in EK~Dra compared with $\alpha$~Cen.  This is a clear indication of enhanced electron densities in the more active star.  The $\alpha$~Cen ratios fall near the low-density limits of the respective $R$'s, indicating $n_{\rm e}$ just below $10^{10}$~cm$^{-3}$; while the EK~Dra ratios both are consistent with densities in the middle range, near or above $10^{11}$~cm$^{-3}$, an order of magnitude larger.  Curiously, the inferred densities are not much different between the more quiet periods of the EK~Dra COS campaign and the large FUV flare decay at the beginning; especially for $R_{2}\ast$, which has the largest grasp of the two diagnostic ratios (i.e., largest variation in $R$ over the range of density sensitivity).

The \ion{O}{4}] densities inferred here are about a factor of 2 higher than those estimated by G\"udel et al.\ (1995) from emission-measure modeling of the {\em ROSAT}\/ survey pulse-height spectra, at least for the softer (2~MK) component of their 2-temperature fit (this is the component that showed rotational modulations, as mentioned earlier; whereas the harder, more dominant, $\sim 10$~MK one showed essentially no phase-correlated variability).  Nevertheless, the agreement must be tempered by the fact that the subcoronal gas is only one-tenth the temperature of even the cooler coronal plasma component, and likely is not co-located.

The high densities during the flare decay in visit~2 suggest short cooling times ($t\sim (n_{\rm e})^{-1}$), at least in the hot-line temperature range ($\sim 10^{5}$~K) where the radiative cooling curve peaks.  This seemingly conflicts with the hours-long progression of the FUV decay.  A similar dichotomy confronts interpretations of long-duration flares on the Sun and other late-type stars.  The usual explanation is that the long duration events are analogous to two-ribbon (2-R) flares on the Sun (e.g., G\"udel 2004), which involve a sequential cascade of excitation through large arcade-like magnetic structures, rather than an impulsive burst in a single coronal loop.  Such 2-R events tend to be quite hot, exceeding 10~MK in peak temperature, with coronal electron densities up to about $10^{11}$~cm$^{-3}$ (e.g., G\"udel 2004).  Importantly, the decay of the flare plasma in a 2-R event can be dominated by conductive energy exchange with the lower cooler layers, rather than direct radiative losses at the super-hot coronal temperatures (see Aschwanden et al.\ 2008).  Thus, the bright FUV emissions and subsonic plasma draining seen in the hot lines of EK~Dra during the large FUV flare decay likely are a byproduct of these cooling processes playing out in the much more energetic setting of the hyper-active young solar analog.  This brings to mind, as intimated by Ayres \& France, the dynamic ``coronal rain'' phenomenon associated with post-flare loops on the Sun, as well as solar active regions in general (see, e.g., Antolin \& Rouppe van der Voort 2012, and references therein).

\subsection{The EK~Draconis Event in a Broader Context}

Because of the unique nature of the time-resolved FUV spectroscopy of the serendipitous EK~Dra flare decay, it is important to place the event within the context of the nearly 100 large stellar flares captured in the soft X-ray band by historical and contemporary high-energy observatories (as documented in the review by G\"udel 2004).  Aschwanden et al.\ (2008) have fitted various combinations of the derived emission measures (EM), X-ray luminosities ($L_{\rm X}$) at the flare temperature peak ($T_{\rm p}$), and decay times ($t_{\rm f}$) with scaling relations approximated as power laws.  Inverting their eq.~12 for the dependence of decay time on flare peak temperature yields an estimate of the (here unknown) $T_{\rm p}$ given the (known) $t_{\rm f}$:
\begin{equation}
T_{\rm p}\sim 10~{\rm MK}\,(\frac{t_{\rm f}}{300~{\rm s}})^{0.6}\,\,\, .
\end{equation}
Given the apparent, albeit FUV, flare decay time of about 7~ks, the peak X-ray temperature would be about 60~MK.  Using this value in eq.~13 of Aschwanden et al.\ yields $L_{X}\sim 10^{31.3}$ erg s$^{-1}$, which is an order of magnitude larger than the ``quiescent'' $L_{\rm X}$ inferred from scaling the contemporaneous [\ion{Fe}{21}] flux from the non-flare intervals.  Supporting that estimate is the fact that the peak enhancement of \ion{C}{4} seen in COS visit~2 is about a factor of 10, and the [\ion{Fe}{21}] proxy for the high-temperature soft X-ray flux -- although recorded later in visit~2, not simultaneously with \ion{C}{4} -- parallels the \ion{C}{4} trajectory during the flare decay, suggesting that [\ion{Fe}{21}] would have been enhanced by a similar large factor, and so too the soft X-ray flux.

The inferred values of flare peak temperature, X-ray luminosity, decay time, and density (again with the caveat that the latter is from the cooler FUV plasma) place the EK~Dra event in the upper-middle range of the documented stellar X-ray flares, as summarized by Aschwanden et al.; perhaps typical of a young active sunlike star, but more accessible in the case of EK~Dra by its proximity: several times closer to Earth than the members of the similar-age young clusters Pleiades and Alpha Persei.

\section{CONCLUSIONS}

The joint STIS/COS study of EK~Dra was adversely impacted by the unfortuitously-timed large FUV flare (interrupting the intended instrumental cross-calibration), but nevertheless the event itself provided valuable insight into the post-flare cooling process, mainly thanks to the high spectral resolution achieved for a variety of key chromospheric and higher temperature features.  The apparent quiescent periods following the flare also contributed significant insight, not only with regard to the velocity cross-calibration of the two UV spectrographs, but also in the lack of phase-correlated profile variations caused by rotational modulations.

To summarize: (1) the FUV redshifts appear to be a characteristic of the normal surface activity of the star, rather than an accident of Doppler Imaging; (2) the hot-line profiles did not change qualitatively during the flare outburst (ten-times enhancement of \ion{C}{4}), except for the larger redshifts of the broad components; and (3) \ion{O}{4}]-derived densities were not very different between the quiescent periods and the FUV flare, despite the fact that the high densities themselves are characteristic of large stellar soft X-ray events (see, e.g., Benz \& G\"udel 2010).

All of these properties suggest that the normal surface conditions on the star are not far removed from those in a big flare; in other words, the stellar corona might be dominated by flaring structures, with a distribution weighted toward the smaller events, as on the Sun, giving the appearance of quiescence at times, when exactly the opposite is true (see G\"udel 2004 for a review of the ``flares all the way down'' coronal scenario; what might be described as a {\em ``flare-ona''}\,).

One notable outcome of the time-resolved flare spectroscopy was the lack of strongly blueshifted emission, which naively might be anticipated from the catastrophic energy release, and indeed such outbursts on the Sun are violently eruptive.  However, as has been noted often in the flare literature, during the explosive heating of the lower atmosphere in an event, most of the outwardly expanding material undoubtedly is too hot to be visible in the FUV lines.  Only the subsequent decay phase would contribute strongly to the FUV features, as the super-hot coronal plasma in the flaring arcades conductively redistributes heat into the lower atmosphere, where it is radiated away partially at FUV temperatures and inspires the subsonic plasma draining phenomenon manifested in the hot-line redshifts.  Unfortunately, the initial impulsive phase of the EK~Dra flare was not captured, and the key diagnostic of the hot-up/cool-down scenario -- [\ion{Fe}{21}] -- was recorded only during the later phases of the event decay (although note in Fig.~4b the apparent small blueshifts of the coronal forbidden line compared with the average visits 3-5 behavior, which would be more exaggerated with the Young et al.\ reference wavelength).

It also must be acknowledged that the peculiarities of the EK~Dra flare decay should be viewed with some caution, because rarely has such an event been recorded in as much detail as provided by the COS part of the {\em HST}\/ campaign: the oddities of the emission Doppler shifts might well be associated with an unusual spatial orientation of the event, for example.

In the end, highly exaggerated stellar incidents like the EK~Dra FUV flare dutifully cause us to look back to the Sun in an effort to identify associations with solar high-energy transient phenomena, such as the coronal rain mentioned earlier.   Nevertheless, drawing such conclusions in general is tricky given the rather tepid activity levels displayed by our local star, even at its worst.  Probably a good thing for habitability of Planet Earth, but a continual challenge for the solar-stellar connection.

\acknowledgments
This work was supported by grant GO-12566 from the Space Telescope Science Institute, based on observations from {\em Hubble Space Telescope}\/ collected at STScI, operated by the Associated Universities for Research in Astronomy, under contract to NASA.  This research also made use of public databases hosted by {SIMBAD}, maintained by {CDS}, Strasbourg, France.



\begin{thebibliography}{}
\bibitem[Antolin \& Rouppe van der Voort(2012)]{2012ApJ...745..152A} Antolin, P., \& Rouppe van der Voort, L.\ 2012, \apj, 745, 152
\bibitem[Aschwanden et al.(2008)]{2008ApJ...672..659A} Aschwanden, M.~J., Stern, R.~A., \& G\"udel, M.\ 2008, \apj, 672, 659
\bibitem[Ayres(2015)]{2015AJ....149...58A} Ayres, T.~R.\ 2015, \aj, 149, 58
\bibitem[Ayres et al.(2003)]{2003ApJ...583..963A} Ayres, T.~R., Brown, A., Harper, G.~M., et al.\ 2003, \apj, 583, 963
\bibitem[Ayres et al.(2005)]{2005ApJ...627L..53A} Ayres, T.~R., Brown, A., \& Harper, G.~M.\ 2005, \apjl, 627, L53
\bibitem[Ayres et al.(1995)]{1995ApJS...96..223A} Ayres, T.~R., Fleming, T.~A., Simon, T.\ et al.\ 1995, \apjs, 96, 223
\bibitem[Ayres \& France(2010)]{2010ApJ...723L..38A} Ayres, T., \& France, K.\ 2010, \apjl, 723, L38
\bibitem[Benz \& G\"udel(2010)]{2010ARA&A..48..241B} Benz, A.~O., \& G\"udel, M.\ 2010, \araa, 48, 241
\bibitem[Candelaresi et al.(2014)]{2014ApJ...792...67C} Candelaresi, S., Hillier, A., Maehara, H., Brandenburg, A., \& Shibata, K.\ 2014, \apj, 792, 67
\bibitem[Cook et al.(1995)]{1995ApJ...444..936C} Cook, J.~W., Keenan, F.~P., Dufton, P.~L., et al.\ 1995, \apj, 444, 936
\bibitem[Drake et al.(2014)]{2014ATel.6121....1D} Drake, S., Osten, R., Page, K.~L., et al.\ 2014, The Astronomer's Telegram, 6121, 1
\bibitem[Fang \& Ding(1995)]{1995A&AS..110...99F} Fang, C., \& Ding, M.~D.\ 1995, \aaps, 110, 99
\bibitem[Fender et al.(2015)]{2015MNRAS.446L..66F} Fender, R.~P., Anderson, G.~E., Osten, R., et al.\ 2015, \mnras, 446, L66
\bibitem[Flower \& Nussbaumer(1975)]{1975A&A....45..145F} Flower, D.~R., \& Nussbaumer, H.\ 1975, \aap, 45, 145 
\bibitem[Green et al.(2012)]{2012ApJ...744...60G} Green, J.~C., Froning, C.~S., Osterman, S., et al.\ 2012, \apj, 744, 60
\bibitem[Gry \& Jenkins(2014)]{2014A&A...567A..58G} Gry, C., \& Jenkins, E.~B.\ 2014, \aap, 567, 58
\bibitem[G\"udel(2004)]{2004A&ARv..12...71G} G\"udel, M.\ 2004, \aapr, 12, 71
\bibitem[G\"udel et al.(1995)]{1995A&A...301..201G} G\"udel, M., Schmitt, J.~H.~M.~M., Benz, A.~O., \& Elias, N.~M., II 1995, \aap, 301, 201
\bibitem[Hawley et al.(2003)]{2003ApJ...597..535H} Hawley, S.~L., Allred, J.~C., Johns-Krull, C.~M., et al.\ 2003, \apj, 597, 535
\bibitem[Keenan et al.(2009)]{2009A&A...495..359K} Keenan, F.~P., Crockett, P.~J., Aggarwal, K.~M., Jess, D.~B., \& Mathioudakis, M.\ 2009, \aap, 495, 359
\bibitem[Kimble et al.(1998)]{1998ApJ...492L..83K} Kimble, R.~A., Woodgate, B.~E., Bowers, C.~W., et al.\ 1998, \apjl, 492, L83
\bibitem[K{\"o}nig et al.(2005)]{2005A&A...435..215K} K{\"o}nig, B., Guenther, E.~W., Woitas, J., \& Hatzes, A.~P.\ 2005, \aap, 435, 215
\bibitem[Kramida et al.(2013)]{} Kramida, A., Ralchenko, Yu., Reader, J., and NIST ASD Team 2013, NIST Atomic Spectra Database (ver.\ 5.1), [Online]. Available: http://physics.nist.gov/asd [2014, August 21]. National Institute of Standards and Technology, Gaithersburg, MD.
\bibitem[Maehara et al.(2012)]{2012Natur.485..478M} Maehara, H., Shibayama, T., Notsu, S., et al.\ 2012, \nat, 485, 478
\bibitem[Mitra-Kraev et al.(2005)]{2005A&A...431..679M} Mitra-Kraev, U., Harra, L.~K., G{\"u}del, M., et al.\ 2005, \aap, 431, 679
\bibitem[Osten et al.(2010)]{2010ApJ...721..785O} Osten, R.~A., Godet, O., Drake, S., et al.\ 2010, \apj, 721, 785
\bibitem[Peter(2006)]{2006A&A...449..759P} Peter, H.\ 2006, \aap, 449, 759
\bibitem[Redfield \& Linsky(2008)]{2008ApJ...673..283R} Redfield, S., \& Linsky, J.~L.\ 2008, \apj, 673, 283
\bibitem[Robinson et al.(2001)]{2001ApJ...554..368R} Robinson, R.~D., Linsky, J.~L., Woodgate, B.~E., \& Timothy, J.~G.\ 2001, \apj, 554, 368
\bibitem[Saar \& Bookbinder(1998)]{1998ASPC..154.1560S} Saar, S.~H., \& Bookbinder, J.~A.\ 1998, in Tenth Cambridge Workshop on Cool Stars, Stellar Systems, and the Sun, (Eds) R. A. Donahue and J. A. Bookbinder, ASP Conf.\ Ser.\ 154, 1560
\bibitem[Schaefer(1989)]{1989ApJ...337..927S} Schaefer, B.~E.\ 1989, \apj, 337, 927 
\bibitem[Schaefer et al.(2000)]{2000ApJ...529.1026S} Schaefer, B.~E., King, J.~R., \& Deliyannis, C.~P.\ 2000, \apj, 529, 1026
\bibitem[Shibayama et al.(2013)]{2013ApJS..209....5S} Shibayama, T., Maehara, H., Notsu, S., et al.\ 2013, \apjs, 209, 5
\bibitem[Tsang et al.(2012)]{2012ApJ...754..107T} Tsang, B.~T.~H., Pun, C.~S.~J., Di Stefano, R., Li, K.~L., 
\& Kong, A.~K.~H.\ 2012, \apj, 754, 107
\bibitem[Woodgate et al.(1998)]{1998PASP..110.1183W} Woodgate, B.~E., Kimble, R.~A., Bowers, C.~W., et al.\ 1998, \pasp, 110, 1183
\bibitem[Young et al.(2015)]{2015ApJ...799..218Y} Young, P.~R., Tian, H., \& Jaeggli, S.\ 2015, \apj, 799, 218 
\end{thebibliography}
\end{document}